\documentclass[a4paper,11pt]{article}

\usepackage{natbib}
\usepackage{eqnarray,amsmath}
\usepackage{amssymb}
\usepackage{graphicx}
\usepackage{epstopdf}
\usepackage{algorithm}
\usepackage{algpseudocode}
\usepackage{subfigure}
\usepackage{bm}
\usepackage[margin=3cm]{geometry}
\usepackage{url}

\def\y{\mathbf{y}}
\def \zb{\mathbf{z}}

\def \E{\mathbb{E}}

\newcommand{\dt}{\text{d}}

\mathchardef\mhyphen="2D 

\title{\bf A Comparison of Likelihood-Free Methods With and Without Summary Statistics}

\author{Christopher Drovandi$^{\dagger, \ddagger, \P,}$\footnote{email: c.drovandi@qut.edu.au} and David T. Frazier$^{\#,\P}$ \\
	\\
	$^\dagger$ School of Mathematical Sciences, Queensland University of Technology (QUT)\\
	$\ddagger$ QUT Centre for Data Science \\
	$\#$ Department of Econometrics and Business Statistics, Monash University \\
	$\P$ Australian Research Council Centre of Excellence for Mathematical and Statistical Frontiers\\
}

\begin{document}

\newcommand{\vect}[1]{\boldsymbol{#1}}

\setlength{\parindent}{0pc}
\setlength{\parskip}{1ex}

\maketitle
\begin{abstract}
	
Likelihood-free methods are useful for parameter estimation of complex models with intractable likelihood functions for which it is easy to simulate data. Such models are prevalent in many disciplines including genetics, biology, ecology and cosmology. Likelihood-free methods avoid explicit likelihood evaluation by finding parameter values of the model that generate data close to the observed data. The general consensus has been that it is most efficient to compare datasets on the basis of a low dimensional informative summary statistic, incurring information loss in favour of reduced dimensionality. More recently, researchers have explored various approaches for efficiently comparing empirical distributions of the data in the likelihood-free context in an effort to avoid data summarisation. This article provides a review of these full data distance based approaches, and conducts the first comprehensive comparison of such methods, both qualitatively and empirically.  We also conduct a substantive empirical comparison with summary statistic based likelihood-free methods.  The discussion and results offer guidance to practitioners considering a likelihood-free approach.  Whilst we find the best approach to be problem dependent, we also find that the full data distance based approaches are promising and warrant further development. We discuss some opportunities for future research in this space.  Computer code to implement the methods discussed in this paper can be found at \url{https://github.com/cdrovandi/ABC-dist-compare}.

\end{abstract}
\noindent
{\it Keywords: approximate Bayesian computation, Bayesian synthetic likelihood, distance function, divergence, generative models, implicit models} 

\newpage

\section{Introduction}
\label{sec:intro}

Likelihood-free Bayesian statistical inference methods are now commonly applied in many different fields. The appeal of such methods is that they do not require a tractable expression for the likelihood function of the proposed model, only the ability to simulate from it.  In essence, proposed parameter values of the model are retained if they produce simulated data `close enough' to the observed data. This gives practitioners great flexibility in designing complex models that more closely resemble reality.

Two popular methods for likelihood-free Bayesian inference that have received considerable attention in the statistical literature are approximate Bayesian computation (ABC, \citet{sisson2018}) and Bayesian synthetic likelihood (BSL, \citet{price2018bayesian,Wood2010}). These approaches traditionally assess the `closeness' of observed and simulated data on the basis of a set of summary statistics believed to be informative about the parameters.  Both ABC and BSL approximate the likelihood of the observed summary statistic via model simulation, but their estimators take different forms.  ABC effectively uses a non-parametric estimate of the summary statistic likelihood \citep{Blum2009}, while BSL uses a parametric approximation via a Gaussian density.

In the context of ABC, the use of a reasonably low-dimensional summary statistic was often seen as necessary to avoid the curse of dimensionality associated with nonparametric conditional density estimation (see \citealp{Blum2009} for a discussion of this phenomena in the context of ABC).  The intuition is that it is difficult to assess closeness of high dimensional datasets in Euclidean space.  Due to its parametric nature, provided that the distribution of the model summary statistic is sufficiently regular, BSL can cope with a higher dimensional summary statistic relative to ABC \citep{price2018bayesian}.  However, BSL ultimately suffers from the same curse.

Recently, there has been a surge of likelihood-free literature that challenge the need for data reduction.  The appeal of such approaches is two-fold: firstly, these approaches bypass the difficult issue of selecting useful summary statistics, which are often model and application specific; secondly, depending on the method, and in the limit of infinite computational resources, it may be feasible to recover the exact posterior. The latter is typically not true of summary statistic based approaches, since in almost all cases the statistic is not sufficient (i.e.\ a loss of information).  The ultimate question is whether full data approaches can mitigate the curse of dimensionality enough to outperform data reduction approaches.  This paper aims to provide insights into the answer to that question.

In the context of ABC, several distance functions have been proposed that compare full observed and simulated datasets via their empirical distributions. For example, the following have been considered: maximum mean discrepancy \citep{Park2016}, Kullback-Leibler divergence (\citealp{jiang2018approximate}), Wasserstein distance \citep{Bernton2019}, energy distance \citep{Nguyen2020}, Hellinger distance \citep{Frazier2020}, and the Cramer von Mises distance \citep{Frazier2020}. Furthermore, in the case of independent observations, \citet{Turner2014} propose an alternative likelihood-free estimator that uses kernel density estimation.  However, while there has been some comparison between the different methods, no systematic and comprehensive comparison between the full data approaches and summary statistic based approaches has been undertaken.

This paper has two key contributions.  The first provides a review of full data likelihood-free Bayesian methods.  The second provides a comprehensive empirical comparison between full data and summary statistic based approaches.

The paper is outlined as follows.  In Section \ref{sec:likelihood-free} we provide an overview of likelihood-free methods that use summary statistics, focussing on ABC and BSL.  Section \ref{sec:full-data} reviews full data approaches to likelihood-free inference, discusses connections and provides a qualitative comparison of them.  Both the full data and summary statistic based likelihood-free approaches are compared on several examples in Section \ref{sec:examples}.  The examples differ in complexity and we consider both simulated and real data scenarios.  Finally, Section \ref{sec:discussion} concludes the paper with a discussion and outlines directions for further research.

\section{Likelihood-Free Bayesian Inference}  \label{sec:likelihood-free}

We observe data ${\y}=(y_{1},\dots,y_{n})^{\top}$, $n\geq 1$, with $y_i\in\mathcal{Y}$ for all $i$, and denote by $P^{(n)}_0$ the true distribution of the observed sample $\y$.  In this paper, we assume that each $y_i \in \mathbb{R}$ is a scalar. However, this assumption can be relaxed for the summary statistic based approaches covered in this section and in the case of certain full data approaches.  The true distribution is unknown and instead we consider that the class of probability measures  
$\mathcal{P}:=\{P^{(n)}_{{\theta }}:\theta\in\Theta\subset \mathbb{R}^{d_{\theta }}\}$, for some value of $\theta$, have generated the data, and denote the corresponding conditional density as $p_n(\cdot\mid\theta)$. Given prior beliefs over the unknown parameters in the model $\theta$, represented by the probability measure $\Pi(\theta)$, with its density denoted by $\pi({\theta })$, our aim is to
produce a good approximation of the exact posterior density 
$$
\pi({\theta \mid \y})\propto p_n(\y\mid\theta)\pi({\theta }).
$$ 

In situations where the likelihood is cumbersome to derive or compute, sampling from $\pi({\theta \mid \y})$ can be computationally costly or infeasible. However, so-called likelihood-free methods can still be used to conduct inference on the unknown parameters $\theta$ by simulating data from the model.  The most common implementations of these methods in the statistical literature are approximate Bayesian computation (ABC) and Bayesian synthetic likelihood (BSL).  Both ABC and BSL generally reduce the data down to a vector of summary statistics and then perform posterior inference on the unknown $\theta$, conditional only on this summary statistic.

More formally, let $\eta(\cdot):\mathbb{R}^{n}\rightarrow\mathbb{R}^{d_\eta}$ denote a $d_\eta$-dimensional map, $d_\eta\ge d_\theta$, that represents the chosen summary statistics, and let $\zb:=(z_1,\dots,z_n)^{\top}\sim P_\theta$ denote data simulated from the model $P^{(n)}_\theta$. For $G_n(\cdot\mid\theta)$ denoting the projection of $P^{(n)}_\theta$ under $\eta(\cdot):\mathbb{R}^n\rightarrow\mathbb{R}^{d_\eta}$, with $g_n(\cdot\mid\theta)$ its corresponding density, the goal of approximate Bayesian methods is to generate samples from the approximate or `partial' posterior 
$$
\pi[{\theta \mid \eta(\y)}]\propto g_n[{\eta(\y)\mid\theta }]\pi(\theta).
$$ 
However, given the complexity of the assumed model, $P^{(n)}_\theta$, it is unlikely that the structure of $G_n(\cdot\mid\theta)$ is any more tractable than the original likelihood function $p_n(\y\mid\theta)$.   Likelihood-free methods such as ABC and BSL employ model simulation to stochastically approximate the summary statistic likelihood in various ways.

ABC approximates the likelihood via the following:
$$
g_{\epsilon}[{\eta(\y)\mid\theta }] = \int_{\mathbb{R}^{d_\eta}} K_\epsilon[\rho\{\eta(\y), \eta(\zb)\}] g_n[{\eta(\y)\mid\theta }] d\zb,
$$
where $\rho\{\eta(\y), \eta(\zb)\}$ measures the discrepancy between observed and simulated summaries and $K_\epsilon[\cdot]$ is a kernel that allocates higher weight to smaller $\rho$.  The bandwidth of the kernel, $\epsilon$, is often referred to  as the tolerance in the ABC literature.  The above integral in intractable, but can be estimated unbiasedly by drawing $m$ mock datasets $\zb_1,\ldots,\zb_m \sim P^{(n)}_\theta$ and computing
$$
\hat{g}_\epsilon[{\eta(\y)\mid\theta }] = \frac{1}{m} \sum_{i=1}^m K_\epsilon[\rho\{\eta(\y), \eta(\zb_i)\}].
$$
In the ABC literature, $m$ is commonly taken to be 1 and the kernel weighting function given by the indicator function, $K_\epsilon[\rho\{\eta(\y), \eta(\zb)\}] = \mathbf{I}[\rho\{\eta(\y), \eta(\zb)\} \leq \epsilon]$.  Using arguments from the exact-approximate literature \citep{Andrieu2009}, unbiasedly estimating the ABC likelihood is enough to produce a Bayesian algorithm that samples from the approximate posterior proportional to $g_\epsilon[{\eta(\y)\mid\theta }]\pi(\theta)$.

As is evident from the above integral estimator, ABC non-parametrically estimates the summary statistic likelihood.  In contrast, BSL uses a parametric estimator.  The standard BSL approach approximates $g_n(\cdot\mid\theta)$ using a Gaussian working likelihood
$$
g_A[{\eta(\y)\mid\theta }] = \mathcal{N}\left[\eta(\y);\mu(\theta),\Sigma(\theta)\right],
$$
where $\mu(\theta)$ and $\Sigma(\theta)$ denote the mean and variance of the model summary statistic at $\theta$.   In almost any practical example $\mu(\theta)$ and $\Sigma(\theta)$ are unknown and we must replace these quantities with those estimated from $m$ independent model simulations.  The standard approach is to use the sample mean and variance:
\begin{flalign*}
\mu_m(\theta)&=\frac{1}{m}\sum_{i=1}^{m}\eta(\zb^i), \\
\Sigma_m(\theta)&=\frac{1}{m}\sum_{i=1}^{m}\left[\eta(\zb^i)-\mu_m(\theta)\right]\left[\eta(\zb^i)-\mu_m(\theta)\right]^{\top},
\end{flalign*}and where each simulated data set $\zb^i$, $i=1,\dots,m$, is generated iid from $P^{(n)}_{\theta}$.  The synthetic likelihood is then approximated as 
$$
\hat{g}_A[{\eta(\y)\mid\theta }] = \mathcal{N}\left[\eta(\y);\mu_m(\theta),\Sigma_m(\theta)\right].
$$
Unlike ABC, $\hat{g}_A[{\eta(\y)\mid\theta }]$ is not an unbiased estimator of $g_A[{\eta(\y)\mid\theta }]$.  However, \citet{price2018bayesian} demonstrate empirically that the BSL posterior depends weakly on $m$, provided that $m$ is chosen large enough so that the plug-in synthetic likelihood estimator has a small enough variance to ensure that MCMC mixing is not adversely affected. More generally, \cite{frazier2020BSLasymp} demonstrate that if the summary statistics are sub-Gaussian, then the choice of $m$ is immaterial so long as $m$ diverges as $n$ diverges. \citet{price2018bayesian} also consider an unbiased estimator of the multivariate normal density for use within BSL.  Given the plug-in estimators' simplicity and its weak dependence on $m$, we do not consider the unbiased version here.

There exist a number of extensions to the standard BSL procedure.  For example, \citet{an2020robust} develop a semi-parametric estimator that is more robust to the Gaussian assumption.  Further, \citet{priddle2019efficient} consider a whitening transformation to de-correlate summary statistics combined with a shrinkage estimator of the covariance to reduce the number of model simulations required to precisely estimate the synthetic likelihood. See \citet{Drovandi2018handbook} for some other extensions to BSL.  For the examples in this paper, we find that the standard BSL method is sufficient to illustrate the results.

The Monte Carlo estimates of the likelihood obtained from ABC or BSL replace the intractable likelihood within a Bayesian algorithm to sample the approximate posterior.  Here we use a Metropolis-Hastings Markov chain Monte Carlo (MCMC) algorithm to sample the ABC or BSL target, in order to ensure that most proposed parameter values are proposed in areas of high (approximate) posterior support.  MCMC was first considered as a sampling algorithm for ABC in \citet{Marjoram2003}, whereas \citet{Wood2010,price2018bayesian} develop MCMC for BSL.

Traditionally, the choice of summary statistics in likelihood-free methods such as ABC and BSL has been crucial.  In the context of ABC, it is generally agreed that one should aim for a low dimensional summary statistic that hopefully carries most of the information contained in the full data.  BSL has been shown to be more tolerant to a higher dimensional summary statistic than ABC, provided that the distribution of the model summary statistic is regular enough \citep{price2018bayesian, frazier2019robust, frazier2020BSLasymp}. However, increasing the number of statistics in BSL will still require increasing the number of model simulations for precisely estimating the synthetic likelihood, so care still needs to be taken.  

\citet{Prangle2018} provides a review of different data dimension reduction methods applied in ABC.  These approaches also hold some relevance for BSL.  Ultimately, the optimal choice of summary statistics will be problem dependent.  For the examples in this paper, we either use a summary statistic that has been reported to perform well from the literature, or the approach we now describe. For the types of examples considered in this paper, namely data with independent observations, a reasonable approach to obtaining useful summary statistics is via indirect inference (e.g.\ \citet{gourieroux1993indirect, Drovandi2014}).  In indirect inference, we construct an auxiliary model with a tractable likelihood $p_A(\y\mid\phi)$ that is parameterised by a vector of unknown parameters $\phi$, where $\phi \in \Phi \subset \mathbb{R}^{d_{\phi}}$ with $d_{\phi} \geq d_{\theta}$. The idea is that the auxiliary model is not mechanistic but can still fit the data reasonably well and thus capture its statistical features.  Either the parameter estimate \citep{DrovandiEtAl2011} or the score of the auxiliary model \citep{Gleim} can be used to form the summary statistic.  Here we use the score, since it only requires fitting the auxiliary model to the observed data and not any datasets simulated during ABC or BSL.  For an arbitrary dataset $\zb$, the score function is given by
$$
S(\zb,\phi^*) =\left. \frac{\partial \log p_A(\zb\mid\phi)}{\partial \phi} \right\rvert_{\phi = \phi^*}.
$$    
The observed statistic is then $\eta(\y) = S(\y,\phi(\y))$ where $\phi(\y) = \arg \max_\phi p_A(\y\mid\phi)$ is the maximum likelihood estimate (MLE).  Thus, the observed statistic is a vector of zeros of length $d_{\phi}$.  We drop $\phi(\y)$ from the notation of the summary statistic, since it remains fixed throughout.  That is, we evaluate the score at $\phi(\y)$ for any dataset $\zb$ simulated in ABC or BSL.  For ABC with summary statistics we use the Mahalanobis distance as the discrepancy function.  The weighting matrix of the Mahalanobis distance is $\vect{J}(\vect{\phi}(\vect{y}))^{-1}$, where $\vect{J}(\vect{\phi}(\vect{y}))$ is the observed information matrix evaluated at the observed data and MLE $\phi(\y)$.

A criticism of summary statistic based approaches is that their choice is often \textit{ad hoc} and there will generally be an inherent loss of information, i.e.\ $\pi[{\theta \mid \eta(\y)}] \neq \pi[{\theta \mid \y}]$. Apart from exponential family models, which appear infrequently in the likelihood-free literature, sufficient statistics of dimension lower than the dimension of the full data do not exist.  Indeed, the use of summary statistics has often been considered a necessary evil to overcome the curse of dimensionality of likelihood-free methods.  However, there has recently been a surge of new approaches that seek to avoid summary statistic selection in favour of directly comparing, in an appropriate distance, the observed and simulated samples. By avoiding summarisation via the direct comparison of observed and simulated data, in a well-chosen distance, the hope is that these approaches will yield more informative inference on the unknown parameters.

\section{Full Data Approaches}  \label{sec:full-data}
All approaches to likelihood-free inference discussed so far rely on the explicit use of a summary statistic $\eta(\y)$ that is of much lower dimension than $\y$. The need to consider low dimensional summaries is due to the fact that estimating $\pi[\theta\mid\eta(\y)]$ via commonly applied algorithms is akin to nonparametric conditional density estimation, i.e., estimating the density of $\theta$ conditional on $\eta(\y)$, and it is well-known that the accuracy of nonparametric conditional density estimators degrades rapidly as the dimension of $\eta(\y)$ increases (see \citealp{Blum2009} for an in-depth discussion on this point). On an intuitive level, the curse of dimensionality is caused by the fact that in a high-dimensional Euclidean space, almost all vectors in that space, e.g., $\y$ and $\zb$, are equally as distant from each other, so that discerning differences between any two vectors becomes increasingly difficult as the dimension increases. 

Ultimately, the curse of dimensionality has led to a fundamental tension in likelihood-free inference: researchers must either make use of exorbitant computational resources in order to reliably compare $\y$ and $\zb$, or they must reduce the data down to summaries $\eta(\y)$, which can entail an excessive loss of information if $\eta(\y)$ is not chosen carefully. However, this tension can actually be cut if we move away from attempting to compare elements in Euclidean space, i.e., comparing $\y$ and $\zb$, and instead try to compare $\y$ and $\zb$ using their probability distributions.

From the above observation, several approaches for comparing observed and simulated datasets via their distributions have recently been proposed within ABC inference. Recall that $\mathcal{P}$ denotes the collection of models used to simulated data, $P_\theta^{(n)}\in\mathcal{P}$ denotes the distribution of $\zb\mid\theta$, $P_0^{(n)}\in\mathcal{P}$ the distribution of $\y$, and define $\rho:\mathcal{P}\times\mathcal{P}\rightarrow\mathbb{R}_{+}$ to be a statistical distance on the space of probability distributions $\mathcal{P}$.\footnote{We note that in general only a few of the full data methods proposed in the literature are actual metrics/norms. While this complicates the mathematics surrounding verification of certain theoretical properties, it has not stymied the use of such distances in practice.} 

Likelihood-free methods based on $\rho(\cdot,\cdot)$ seek to select draws of $\theta$ such that $\rho(P^{(n)}_0,P_\theta^{(n)})$ is ``small'' with large probability. This construction means that the likelihood-free methods based on comparing distributions can differ in two ways: one, the choice of $\rho(\cdot,\cdot)$; two, their use of the kernel $K_\epsilon$ in constructing the posterior for $\theta$. Since the second aspect has been shown to be largely immaterial to inference in the case of summary statistic based ABC, in this review we focus on the choice of $\rho(\cdot,\cdot)$. 

In practice, calculating $\rho(P_\theta^{(n)},P_0^{(n)})$ is infeasible since $P_0^{(n)}$ is unknown and $P_\theta^{(n)}$ is intractable. To circumvent this issue, instead of attempting to compare the joint laws $P_0^{(n)}$ and $P_\theta^{(n)}$, existing full data approaches only compare marginal distributions. The latter (marginal) distributions can be conveniently estimated using the empirical distributions of $\y$ and $\zb$: for $\delta_{x}$ denoting the Dirac measure on $x\in\mathcal{Y}$, define the empirical distribution of the observed data $\y$ as $\hat{\mu} = n^{-1}\sum_{i=1}^n \delta_{y_i}$, and, for any $\theta\in\Theta$, let $\hat{\mu}_{\theta}=n^{-1}\sum_{i=1}^n \delta_{z_i}$, where $\zb\sim P_\theta^{(n)}$, denote the empirical distribution of the simulated data. Even though the likelihood is intractable, $\hat\mu$ and $\hat\mu_{\theta}$ can always be constructed. 

Given an observed dataset $\y$, and a particular choice for $\rho(\cdot,\cdot)$, the ABC posterior based on the statistical distance $\rho(\cdot,\cdot)$ uses the  likelihood 
\begin{equation}\label{eq:like_fd}
g^\rho_\epsilon[\y\mid\theta]=\int_{\mathcal{Y}}K_\epsilon\left[\rho\left(\hat\mu,\hat\mu_{\theta}\right)\right]p_n(\zb\mid\theta)\dt\zb,
\end{equation}
and yields the ABC posterior
$$
\pi^\rho_\epsilon[\theta\mid\y]\propto g^\rho_\epsilon[\y\mid\theta]\pi(\theta).
$$The posterior notation $\pi^\rho_\epsilon[\theta\mid\y]$ highlights the fact that this posterior is conditioned on the entire sample of observed data $\y$ (via $\hat\mu$) and depends on the choice of distance $\rho(\cdot,\cdot)$.

While there are many possible distances to choose from, and thus many different posteriors one could compute, two different choices of $\rho(\cdot,\cdot)$ can deliver posteriors that vary significantly from one another. Moreover, the resources necessary to compute the posterior under different choices for $\rho(\cdot,\cdot)$ can also vary drastically. In addition, not all distances on $\mathcal{P}$ are created equal; certain distances yield more reliable posterior approximations than others depending on the size, type, and variability of the data. Given these issues regarding the choice of $\rho(\cdot,\cdot)$, in what follows we review several approaches that have been employed in the literature and attempt to highlight in what types of problems they are best suited.\footnote{In what follows, we note that many of the distances presented can be extended to cases where $\y$ is multivariate, and to cases where $\y$ and $\zb$ are computed using a differing numbers of observations. However, as these issues are not entirely germane to the production of the ABC posterior based on this distance, or the resulting accuracy of the posteriors across different methods, we do not discuss these extensions herein.}

\subsubsection*{Wasserstein Distance.} One of the most commonly employed approaches to full data inference in ABC, as proposed by \cite{Bernton2019}, takes $\rho(\cdot,\cdot)$ to be the Wasserstein distance. Let $(\mathcal{Y},d)$ be a metric space, and for $p\ge1$ let $\mathcal{P}_p(\mathcal{Y})$ denote the collection of all probability measures $\mu$ defined on $\mathcal{Y}$ with finite $p$-th moment. Then, in the case of scalar random variables, the $p$-Wasserstein distance on $\mathcal{P}$ between $\mu,\nu\in\mathcal{P}_p(\mathcal{Y})$ can be defined as
$$
\mathcal{W}_p(\mu,\nu)=\left(\int_{0}^{1}|F^{-1}_\mu(\lambda)-F_\nu^{-1}(\lambda)|^{p}\dt \lambda\right)^{1/p},$$where $F_\mu(\cdot)$ denotes the cumulative distribution function (CDF) of the distribution $\mu$, and $F^{-1}_\mu(\cdot)$ its quantile function. For a review of the Wasserstein distance, and optimal transport more broadly, we refer to \cite{villani2008optimal}.  

While the above formula looks complicated, the Wasserstein distance between the empirical distributions $\hat\mu$ and $\hat\mu_\theta$ takes a simpler form in the case of $p=1$. Namely, for $y_{(i)}$ denoting the $i$-th sample order statistic, 
$$
\mathcal{W}_{1}\left(\hat\mu, \hat\mu_\theta\right)=n^{-1}\sum_{j=1}^{n} |y_{(i)}- z_{(i)}|,
$$
which is nothing but comparing, in the $L_1$ norm, the (average of the) $n$ order statistics calculated from $\y$ and $\zb$. As such, calculation of $\mathcal{W}_{1}\left(\hat\mu, \hat\mu_\theta\right)$ only requires sorting the samples (separately) and taking the absolute difference between the observed and simulated order statistics.

The use of $\mathcal{W}_{1}\left(\hat\mu, \hat\mu_\theta\right)$ within ABC, by replacing $\rho(\hat\mu,\hat\mu_\theta)$ in \eqref{eq:like_fd} by $\mathcal{W}_{1}\left(\hat\mu, \hat\mu_\theta\right)$, can be interpreted as matching \textit{all} quantiles of the empirical and simulated distributions. We  note here that the use of quantiles as summary statistics in ABC is commonplace (see, e.g., \citealp{fearnhead2012constructing}). Given this interpretation, we would expect that ABC based on $\mathcal{W}_{1}\left(\hat\mu, \hat\mu_\theta\right)$ will produce reliable posterior approximations in situations where the quantiles of the distribution are sensitive to fluctuations in $\theta$. However, if the quantiles of $\zb$ do not vary significantly as $\theta$ changes, the Wasserstein distance will not change in a meaningful manner, and the posterior approximation may be poor. For instance, if the data displays dynamic time-varying features in certain conditional moments, then it may be difficult for ABC based on the Wasserstein to account for these features, and the approach may have to be supplemented with additional summaries that specifically target the dynamics inherent in the series. 

Lastly, we note that while ABC based on the Wasserstein is a ``black-box'' approach to choosing summaries, since the approach boils down to matching sorted samples of observations in the $L_1$ norm, we may encounter the curse of dimensionality in situations where $n$ is large.

\subsubsection*{Energy Distance.} Energy distances (or statistics) are classes of functions for measuring the discrepancy between two random variables; we refer to \cite{szekely2005new} for a review of the energy distances and their applications in statistics. To define the energy distance let $Y_1\in\mathcal{Y}$ and $Z_1\in\mathcal{Y}$ denote independent variables with distributions $\mu$ and $\nu$, such that $\int_{\mathcal{Y}}\|y_1\|_p\dt\mu(y_1)<\infty$ and $\int_{\mathcal{Y}}\|z_1\|_p\dt\nu(z_1)<\infty$. Also, let $Y_2$ and $Z_2$ denote random variables with the same distribution as $Y_1$ and $Z_1$, respectively, but independent of $Y_1$ and $Z_1$. For an integer $p\ge1$, the $p$-th energy distance $\mathcal{E}_p(\mu,\nu)$ can be defined as  
$$
\mathcal{E}_p(\mu,\nu)=2\mathbb{E}\|Y_1-Z_1\|_p-\mathbb{E}\|Z_1-Z_2\|_p-\mathbb{E}\|Y_1-Y_2\|_p,
$$and satisfies $\mathcal{E}_p(\mu,\nu)\ge0$, with equality if an only if $\mu=\nu$ (\citealp{szekely2005new}). Using this latter inequality, $\sqrt{\mathcal{E}_p(\mu,\nu)}$ can be viewed as a metric on the space of univariate distribution functions.  

The inequality $\mathcal{E}_p(\mu,\nu)\ge0$ provides motivation for using this distance to measure the discrepancy between probability distributions of two separate samples of observations, and in this way is related to other nonparametric two-sample test statistics (such as the Cramer-von Mises statistic discussed later). The ability of the energy distance to reliably discriminate between two distributions has led \cite{Nguyen2020} to use the energy distance to compare $\y$ and $\zb$ in order to produce an ABC-based posterior for $\theta$. Since $\mathcal{E}_p(\mu,\nu)$ cannot be calculated directly, \cite{Nguyen2020} propose to replace the energy distance by the V-statistic estimator
\begin{align*}
\widehat{\mathcal{E}}_p(\hat\mu,\hat\mu_\theta)&=\frac{2}{n^2} \sum_{i=1}^{n} \sum_{j=1}^{n} \|y_i-z_j\|_p- \\
&\frac{1}{n^{2}} \sum_{i=1}^{n} \sum_{j=1}^{n} \|z_i-z_j\|_p -\\
&\frac{1}{n^{2}} \sum_{i=1}^{n} \sum_{j=1}^{n} \|y_i-y_j\|_p,
\end{align*}
and thus set $\rho(\cdot,\cdot)=\widehat{\mathcal{E}}(\hat\mu,\hat\mu_\theta)$ in equation \eqref{eq:like_fd}.\footnote{We note that \cite{gretton2008kernel} have demonstrated that $\widehat{\mathcal{E}}(\hat\mu,\hat\mu_\theta)\rightarrow_p\mathcal{E}_p(\mu,\nu)$ as $n\rightarrow\infty$.}


The main restrictions on the use of $\mathcal{E}_p$ in ABC-based inference relates to its moment restrictions. Existence of $\mathcal{E}_p(\mu,\nu)$ requires at least a $p$-th moment for both variables under analysis. Such an assumption is violated for heavy tailed data, such as stable distributions, which are a commonly encountered example in the ABC literature. Consequently, if there are outliers in the data, ABC inference predicated on the energy distance may not be accurate. 

In addition, the V-statistic estimator $\widehat{\mathcal{E}}_p(\hat\mu,\hat\mu_\theta)$ generally requires $O(n^2)$ computations. Therefore, in situations where $n$ is large, or if many evaluations of $\mathcal{E}_p(\mu,\nu)$ are required, posterior inference based on $\mathcal{E}_p(\mu,\nu)$ may be time consuming.

\subsubsection*{Maximum Mean Discrepancy.} The energy distance is a specific member of the class of maximum mean discrepancy (MMD) distances between two probability measures. Let $k:\mathcal{Y}\times\mathcal{Y}\rightarrow\mathbb{R}_{}$ be a Mercer kernel function,\footnote{That is, $k(\cdot,\cdot)$ is symmetric, continuous, and is positive-definite, i.e., $\sum_{i=1}\sum_{j=1}k(y_i,y_j)c_ic_j\geq0$ for all finite sequences $y_{1},\dots,y_{n}$ on $\mathcal{Y}$ and all real $c_1,\dots,c_n$.} and let $Y_1\in\mathcal{Y}$ and $Z_1\in\mathcal{Y}$ be distributed according to $\mu$ and $\nu$, respectively, with $Y_2,\;Z_2$ again denoting an iid copy of $Y_1,\;Z_1$. Then the MMD between $\mu$ and $\nu$ is given by 
\begin{align*}
\text{MMD}^2(\mu,\nu)&=\E\left[k(Y_1,Y_2)\right]+\E\left[k(Z_1,Z_2)\right]- \\
&2\E\left[k(Y_1,Z_1)\right].
\end{align*}

The choice of kernel in the MMD determines which features of the probability distributions under analysis one is interested in discriminating against. If the kernel is taken to be polynomial, as in the energy distance, then one is interested in capturing differences in moments between the two distributions. If instead one chooses a class of kernels such as the Gaussian, $\exp\left(-\|y-z\|^2_2/2\sigma\right)$ or Laplace, $\exp\left(-|y-z|_1/\sigma\right)$,  then one attempts to match all moments of the two distributions.\footnote{This latter class of kernels is often called \textit{characteristic}; see \cite{gretton2012kernel} and the references therein for further discussion on the use of specific kernel types in MMD.}

As with the energy distance, direct calculation of MMD is infeasible in cases where $\mu,\nu$ are unknown and/or intractable. However, writing MMD in terms of expectations allows us to consider the following estimator based on $\y$ and $\zb$: 
\begin{align*}
\widehat{\text{MMD}}^2(\hat\mu,\hat\mu_\theta)&=\frac{1}{n(n-1)}\sum_{i=1}^{n}\sum_{j\neq i}k(y_i,y_j) + \\
& \frac{1}{n(n-1)}\sum_{i=1}^{n}\sum_{j\neq i}k(z_i,z_j)- \\
&\frac{2}{n^2}\sum_{i=1}^{n}\sum_{j\neq i}k(y_i,z_j).
\end{align*}

The ability to bypass summary statistics via the MMD in ABC was initially proposed by \cite{Park2016}, and has found subsequent use in several studies. The benefits of MMD are most appreciable in cases where initial summary statistics are hard to construct, or in situations where the structure of the data makes constructing a single set of summary statistics to capture all aspects of the data difficult, such as in dynamic queuing networks (\citealp{ebert2018likelihood}).

The MMD estimator $\widehat{\text{MMD}}^2(\hat\mu,\hat\mu_\theta)$ can be seen as an unbiased U-statistic estimator of the population counterpart. Therefore, $\widehat{\text{MMD}}^2(\hat\mu,\hat\mu_\theta)$ need not be bounded below by zero (i.e., it can take negative values). Given this fact, \cite{Nguyen2020} argue that it is not necessarily suitable as a discrepancy measure for use in generative models.

Unlike the Wasserstein or energy distance, the use of MMD requires an explicit choice of kernel function, and it is currently unclear how the resulting choice affects the accuracy of the posterior approximation. In particular, while it is common to consider a Gaussian kernel, it is unclear whether this choice is preferable in all situations. Moreover, we note that, as in the case of the Energy distance, the choice of kernel in MMD automatically imposes an implicit moment assumptions. Namely, the expectations that define the MMD criterion must exist. Therefore, depending on the kernel choice, MMD may not yield reliable posterior inferences if there are outliers in the data or if the data has heavy tails.

In addition, it is important to point out that the calculation of $\widehat{\text{MMD}}^2(\hat\mu,\hat\mu_\theta)$ requires $O(n^2)$ calculations, which can become time consuming when $n$ is large, and/or when many evaluations of $\widehat{\text{MMD}}^2(\hat\mu,\hat\mu_\theta)$ are required to obtain an accurate posterior approximation.\footnote{While faster estimators for MMD, and other distances, may exist, to ensure a fair comparison across different methods we only consider the most basic, and hence direct, estimators of the distances.}

\subsubsection*{Cramer-von Mises Distance.}
The Cramer distance between the empirical CDF of the observed sample, $\hat\mu$, and a theoretical distribution $\mu_\theta$ is defined as the $L_2$ distance between $\hat\mu$ and $\mu_\theta$:
$$
\int_{\mathcal{Y}} [\hat\mu(y)-\mu_\theta(y)]^2\dt y.
$$However, practical use of the above distance is made difficult by the fact that the distribution of the distance depends on the specific $\mu_\theta$ under hypothesis. To rectify this issue, we integrate the Cramer distance with respect to the hypothesised measure, $\mu_\theta$, to obtain the Cramer-von Mises (CvM) distance
$$
\mathcal{C}(\hat\mu,\mu_\theta):=\int_{\mathcal{Y}} [\hat\mu(y)-\mu_\theta(y)]^2\dt \mu_\theta(y),
$$which has a distribution that, by construction, does not depend on $\mu_\theta$ (see, e.g., \citealp{anderson1962distribution}). 

In the case of ABC, the measure $\mu_\theta$ is intractable, so direct calculation of $\mathcal{C}(\hat\mu,\mu_\theta)$ is infeasible and we can instead employ the following estimator of the CvM distance: for $\widehat{H}(t)=\frac{1}{2}\left[\hat\mu_n(t)+\hat\mu_\theta(t)\right]$
\begin{flalign*}
\widehat{\mathcal{C}}(\hat\mu,\hat\mu_\theta)&:=\frac{n}{2}\int_{\mathcal{Y}}\left[\hat\mu(t)-\hat\mu_\theta(t)\right]^2\dt\widehat{H}(t),
\end{flalign*}where $\hat\mu_\theta(t)$ denotes the empirical CDF at the point $t$ based on the simulated data $\mathbf{z}$.\footnote{If the observed sample has $n$ observations, and the simulated sample $m$, the two-sample CvM statistic is given by (see, equation (2) in \citealp{anderson1962distribution})
	$$
	[(nm)/(n+m)]\int [\hat\mu_{\theta}(t)-\hat\mu_n(t)]^2d \widehat{H}(t),
	$$where $(n+m)\widehat{H}(t)=n\hat\mu_{}(t)+m\hat\mu_{\theta}(t)$. When the samples are the same length, i.e., $n=m$, the statistic simplifies to $\widehat{\mathcal{C}}(\hat\mu,\hat\mu_\theta)$ in the displayed equation. }

For continuously distributed data, $\widehat{\mathcal{C}}(\hat\mu,\hat\mu_\theta)$ can be rewritten in terms of the ranks of the observed and simulated samples. Let $h_{(1)}<\dots<h_{(2n)}$ denote the ordered joint sample $\mathbf{h}=(\y',\zb')'$. Define $r_{(1)}<\dots<r_{(n)}$ as the corresponding ranks in $\mathbf{h}$ associated with $\y$, and likewise let $s_{(1)}< \dots< s_{(n)}$ denoted the ranks in $\mathbf{h}$ associated with $\zb$, then \cite{anderson1962distribution} showed that
\begin{flalign*}
\widehat{\mathcal{C}}(\hat\mu,\hat\mu_\theta)&:=\frac{U}{2n^4}-\frac{4n^2-1}{12(n)},\text{ where } \\
U/n&=\sum_{i=1}^{n}(r_{(i)}-i)^2+\sum_{j=1}^{n}(s_{(j)}-j)^2.
\end{flalign*}The above formula makes clear that calculating the CvM distance is quite simple, as it just involves sorting the entire sample, and calculating the corresponding ranks of $\y$ and $\zb$ in the joint sample, $\mathbf{h}$. 

The CvM-statistic has certain advantages over other possible distance choices. Most notably, the CvM distance is robust to heavy-tailed distributions and outliers. This property has immediate benefits in the realm of ABC, where it is common to encounter stable distributed random variables, which may not have any finite moments. Furthermore, the CvM distance can be used in any situation where the ECDF can be reliably estimated; i.e., it can be reliably implemented for independent, weakly dependent or cross-sectionally dependent data. An additional advantage is that inference based on the CvM distances is often less sensitive to model misspecification than inferences based on other distances. This latter property is what motivates \cite{Frazier2020} to apply the CvM in misspecified generative models.\footnote{\cite{Frazier2020} also proposes the use of the Hellinger distance to deliver robust inferences in ABC in the case of misspecified models. To keep this review to a reasonable length, we do not review this distance herein.}

While useful, computation of the CvM distance essentially boils down to estimating two empirical CDFs, which means that if that the sample size is relatively small, the estimated CvM distance can be noisy and the resulting ABC inference poor. Further to this point,  since the CvM distance is based on the difference of two CDFs, which are bounded on $[0,1]$, differences between the CDFs that only occur in the tails of the data become ``pinched'' and are unlikely to result in a ``large'' value of $\widehat{C}(\hat\mu,\hat\mu_\theta)$. Hence, if there are parameters of the model that explicitly capture behavior in the far tails of the data, but do not impact other features of the distribution, such as skewness or kurtosis, then the CvM may yield inaccurate inferences for these parameters.

In terms of computation, the CvM distance is relatively simple to calculate. However, we note that since the CvM requires sorting the joint sample $\mathbf{h}=(\y',\zb')'$ calculation of $\widehat{\mathcal{C}}(\hat\mu,\hat\mu_\theta)$ in large samples may take longer than $\mathcal{W}_1(\hat\mu,\hat\mu_\theta)$, which only requires sorting the individual samples.

\subsubsection*{Kullback-Leibler Divergence.}

The last class of statistical distances we review are those based on Kullback-Leibler (KL) divergence. Given two iid datasets $\y$ and $\zb$, \cite{jiang2018approximate} propose to conduct posterior inference on $\theta$ by choosing as the distance $\rho(\cdot,\cdot)$ in \eqref{eq:like_fd} the KL divergence between the densities of $\y$ and $\zb$. Assume that $\y$ is generated iid from $\mu$ with density $f_\mu:=\dt\mu/\dt\lambda$, and $\zb$ iid from $\nu$ with density $f_\nu:=\dt\nu/\dt\lambda$, where $\dt\lambda$ denotes a dominating measure. The KL divergence between $f_\mu$ and $f_\nu$ is defined as 
$$
\text{KL}(f_\mu,f_\nu)=\int f_\mu(y)\ln\frac{f_\mu(y)}{f_\nu(y)}\dt y,
$$and is zero if and only if $f_\mu=f_\nu$. 

Similar to the other distances discussed above, calculation of $\text{KL}(f_\mu,f_\nu)$ is infeasible in the ABC context. To this end, given observed data $\y$ and simulated data $\zb$, \cite{jiang2018approximate} estimate $\text{KL}(f_\mu,f_\nu)$ using the 1-nearest neighbour density estimator of the KL divergence presented in \cite{perez2008kullback}: 
$$
\widehat{\text{KL}}(\y,\zb)=\frac{1}{n}\sum_{i=1}^{n}\ln\frac{\min_j\|z_i-y_j\|}{\min_{j\neq i}\|z_i-z_j\|}+\ln \frac{n}{n-1}
.$$
The above discrepancy is simple to calculate and has a time cost of $O(n\ln n)$ and thus is only marginally slower to calculate than any of the other distances discuss above, save for the MMD or energy distance, which both have a cost of $O(n^2)$. 

Using $\widehat{\text{KL}}(\y,\zb)$ in \eqref{eq:like_fd}, \cite{jiang2018approximate} compares this approach against other ABC approaches based on both full data distances, such as the Wasserstein, and based on automatic summary statistics \textit{a la} \cite{fearnhead2012constructing}. The results suggest that ABC-inference based on the KL divergence can outperform other measures when the model is correctly specified, and when the data is iid, at least in relatively small samples.\footnote{Across all the simulated examples considered in \cite{jiang2018approximate} the sample size used for analysis was no greater than $n=500$.} 

While useful, the approach of \citet{jiang2018approximate} is only valid for absolutely continuous distributions, and is not applicable for discrete or mixed data. In such cases, and if one still wishes to use something like the KL divergence to conduct ABC, one can instead use the approach proposed by \cite{Turner2014}. 

While the approach of \cite{jiang2018approximate} approximates the KL divergence directly, the approach of \cite{Turner2014} essentially constructs a simulated estimator of the likelihood, for every proposed value of $\theta$, and then evaluates the observed sample at this likelihood estimate. As such, this approach is not strictly speaking an ABC approach, but remains a likelihood-free method. 

To present the approach of \cite{Turner2014}, for simplicity let us focus on the case where $z_i$ is generated from a continuous distribution.\footnote{The case of discrete or mixed data can be handled by considering a kernel density estimator that is appropriate for these settings, and by sufficiently modifying the simulated estimator of the likelihood function.} Then, the approach of \cite{Turner2014} first generates $j=1,\dots,m$ iid realizations of $\zb^{j}=(z_1^j,\dots,z_n^j)'$, with $z^j_i\stackrel{iid}{\sim} P_\theta$, for each $i$ and $j$, and constructs an estimator of the model density at the point $z^\star$ by averaging, over the $m$ datasets, the standard kernel density estimator 
\begin{align*}
\hat{f}_{m,\delta}(z^\star\mid\theta)&=\frac{1}{m}\sum_{j=1}^{m}\hat{f}_\delta(z^\star\mid\zb^{j}),\text{ where } \\
\hat{f}_\delta(z^\star\mid\zb)&=\frac{1}{n\delta}\sum_{i=1}^{n}K_\delta(z^\star-z^j_i),
\end{align*}
where $K_\delta$ is a kernel function with bandwidth parameter $\delta$. Using this density estimator, \cite{Turner2014} construct the estimated likelihood $
\hat{p}_n(\y\mid\theta)=\prod_{i=1}^{n}\hat{f}_{m,\delta}(y_i\mid\theta)
$, and subsequently use $\hat{p}_n(\y\mid\theta)$ in place of the actual likelihood within a given MCMC scheme to conduct posterior inference on $\theta$.  {When the data are iid there is a computational cost saving that can be achieved.  Here the $n \times m$ individual simulated data points can be concatenated into a vector to construct a single kernel density estimate, which is then evaluated at each of the $n$ observed data points. That is, simulated data for observation $i$ can be recycled for observation $k \neq i$.  This may reduce the value of $m$ required.  Indeed, in this iid setting, the size of the single concatenated simulated dataset (here $n \times m$) need not be an exact multiple of $n$.}

The approach of \cite{Turner2014} is not based on a distance between the simulated and observed samples, but on a (simulation-based) estimate of the likelihood. Therefore, in the limit of infinite computational resources, i.e., as $m\rightarrow\infty$, the approach of \cite{Turner2014} will yield the exact likelihood function $p_n(\y\mid\theta)$, and thus the `exact' posterior $\pi(\theta\mid \y)$.\footnote{This convergence can be seen by recalling that, for $p(y\mid\theta)$ denoting the density of $y$ conditional on a value of $\theta\in\Theta$, under regularity conditions, the integrated mean squared difference satisfies $\int\{\hat{f}_{m,\delta}(y\mid\theta)-p(y\mid\theta)\}^2\dt y=O(\delta^4+(nm\delta)^{-1})$ (see, e.g., Theorem 24.1 in \citealp{van2000asymptotic}). Considering $n$ as fixed, each individual likelihood term then converges to $p(y\mid\theta)$ as $m\rightarrow\infty$, so long as $\delta\rightarrow0$ and $m\delta\rightarrow\infty$.} In contrast, even in the case where it is feasible to set the ABC tolerance as $\epsilon=0$, ABC based on the statistical distance $\rho(\cdot,\cdot)$ will only ever deliver an approximation to the `exact' posterior that is particular to the choice of statistical distance. The obvious exception to this statement is the case where the information contained in the statistical distance coincides with that contained in the likelihood (i.e., the Fisher information), which is not generally the case for any of the methods discussed above.

Unlike the distance estimators discussed previously, the approach of \cite{jiang2018approximate} requires iid data, while the approach of \cite{Turner2014} requires (at least) independent data, with the latter approach also requiring additional modifications depending on the model under analysis. Moreover, it is not immediately obvious how to extend these approaches to capture other dependence regimes. Therefore, while these approaches may yield accurate posterior approximations in settings where the density of the model is intractable, e.g., in stable or $g$-and-$k$ distributions, these methods are not appropriate for conducting inference in models with cross-sectional or temporal dependence. In addition, since these approaches are akin to using a simulation-based estimate of the likelihood function (or a function thereof) as a distance, in cases where the model is misspecified, these approaches may perform poorly and alternative measures may yield more reliable inference (see, e.g., the robust BSL approach of \citealp{frazier2019robust}, or the robust ABC approaches discussed in \citealp{Frazier2020}).

\section{Examples} \label{sec:examples}

For the examples we use a Gaussian mixture, parameterised by the component means, standard deviations and weights, as the auxiliary model for forming the summary statistics. Specifically, we use the score function of the Gaussian mixture evaluated at the maximum likelihood estimate (MLE) based on the observed data as the summary statistics. The observed summary statistic, obtained from substituting the observed data into the score function, is thus theoretically equal to a vector of zeros provided that the MLE lies in the interior of the parameter space.  The MLE is obtained using the EM algorithm with multiple random initialisations.  The number of components in the mixture are specified in each example.  In some cases we use different summary statistics, which we define when needed.  In the results (shown as tables and figures), we refer to ABC with summary statistics as simply ABC.   BSL only uses summary statistics, and so we refer to that as BSL.  

When using summary statistics, we use the same ones for both ABC and BSL.  We note that in practice the principles for choosing summaries may differ for ABC and BSL, since BSL is more tolerant to high-dimensional summaries but requires that the distribution of the simulated summaries is reasonably well behaved \citep{frazier2020BSLasymp}. In this paper we choose summaries that are reasonable for both ABC and BSL, in the sense they are low-dimensional and are approximately Gaussian in large samples, in order to more easily compare the methods' performance.

The approach of \citet{Turner2014} that uses kernel density estimation is referred to as KDE in the results.  For the full data distance ABC approaches, we consider the: Cramer von Mises distance (CvM), Wasserstein distance (Wass) and the maximum mean discrepancy (MMD).  For MMD we use a Gaussian kernel with a bandwidth that is set as the median of the Euclidean distances between pairs of data points of the observed data, consistent with \citet{Park2016}.  For KDE, we use a Gaussian kernel with a bandwidth given by Silverman's rule of thumb.

We have chosen the specific distances to use in the following examples based on computational cost and diversity across the methods. In particular, since the energy statistic is a specific member of the MMD family, and since both require $O(n^2)$ computations for a single evaluation, it is prohibitively difficult to consider repeated sampling comparisons using both methods. In addition, the KDE approach and the KL divergence approach have a similar flavour, both can be seen as based on estimated densities, and both are applicable in the same types of settings (i.e., both require independent data). Therefore, to render the comparison between the various methods more computationally feasible, we only consider the KDE approach in what follows.

We use MCMC to sample the approximate posteriors. When parameters are bounded, we use an appropriate logistic transformation to sample an unbounded space.  We use a multivariate normal random walk with a covariance set at an estimate of the relevant approximate posterior obtained from pilot runs.  The number of MCMC iterations is set large to ensure that the Monte Carlo error has little impact on the conclusions drawn. 

For BSL, we choose $m$ so that the standard deviation of the log-likelihood at a central parameter value (true value when available) is roughly between 1 and 2.  For ABC, we take $\epsilon$ as a particular sample quantile of 100K independent simulated ABC discrepancy values based on a central parameter value.  We choose the quantile so that the effective sample size of the MCMC is of the same order as that for BSL for the same total number of model simulations.     In some examples, the overall ABC distance function is a linear combination of multiple distance functions.  For the weights we compute the inverse of the sample standard deviation of the individual discrepancies, or a robust measure thereof when outlier distances are present.   For the ABC approaches (both summary statistic and full data) we always use $m = 1$.  It is less clear how to choose $m$ for KDE compared to BSL, as we find that the posterior based on KDE can be quite sensitive to $m$.  Thus, for KDE, we choose $m$ as large as possible so that the overall efficiency (effective sample size divided by the number of model simulations) remains similar to that of BSL.    Thus, we allocate roughly the same computational effort in terms of the number of model simulations to all approaches.  It is important to note, however, that there can be significant overhead associated with some of the methods.  For example, MMD is slow for larger datasets and KDE involves kernel density estimation, which can be slow when there are a large number of simulated data points used to construct the KDE.  This aspect is discussed further in each of the examples.  

The first two examples are toy and it is possible to compare methods on repeated simulated datasets.  The second two examples are more substantive and computationally intensive, hence we compare methods on one single and real dataset only. For a single simulated dataset, we compare methods visually on the basis of which posterior approximation is more concentrated around the true parameter, since ABC and BSL do not tend to over-concentrate due to the use of summary statistics and/or ABC threshold.  For the single real dataset where the true parameter is not available we consider the concentration of the posterior approximations, guided by the results for the corresponding simulated dataset.

\subsection{g-and-k Example}

The g-and-k distribution (e.g.\ \citet{Rayner2002}) is a complex distribution defined in terms of its quantile function that is commonly used as an illustrative example in likelihood-free research (for early ABC treatments see \citet{Allingham,Drovandi2011}). The quantile function for the g-and-k model is given by
\begin{align}
Q(z(p);{\theta}) &= a + b\left(1 + c\frac{1-\exp(-gz(p))}{1 + \exp(-gz(p))}\right) \nonumber \\
&(1 + z(p)^2)^kz(p). \label{eq:g-and-k}
\end{align}
Here $p$ denotes the quantile of interest while $z(p)$ represents the quantile function of the standard normal distribution.  The model parameter is ${\theta} = (a,b,c,g,k)$, though common practice is to fix $c$ at 0.8, which we do here (see \citet{Rayner2002} for a justification).  The example is suitable to examine the performance of likelihood-free methods since the likelihood can be computed numerically \citep{Rayner2002} permitting exact Bayesian inference, albeit more cumbersome than simulating the model which can be done straightforwardly via inversion sampling. 

\begin{figure}
	\centering
	\includegraphics[scale=0.3]{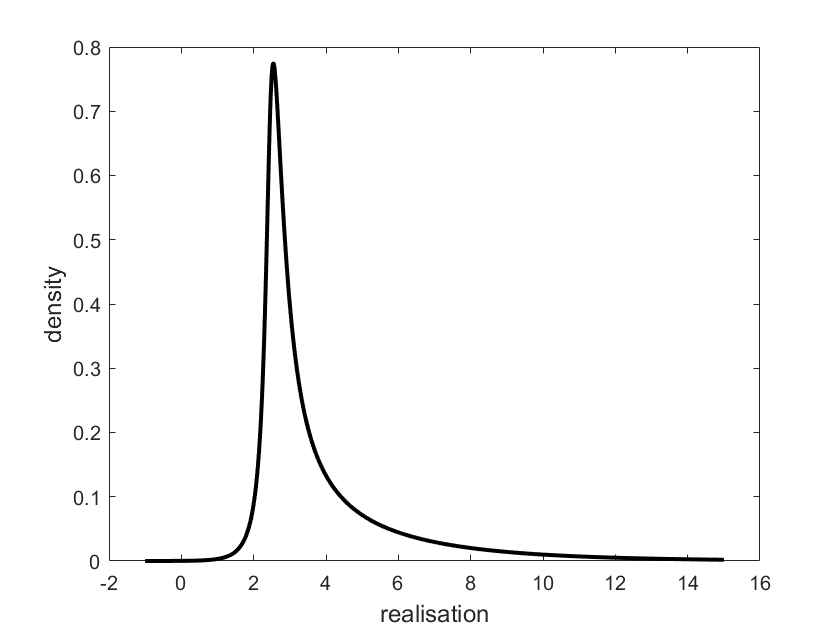}
	\caption{True density of the g-and-k distribution with parameter value $a = 3$, $b = 1$, $g = 2$ and $k = 0.5$.}
	\label{fig:gandk_density}
\end{figure}

Here we consider sample sizes of $n = 100$ and $n = 1000$, with true parameter value $a = 3$, $b = 1$, $g = 2$ and $k = 0.5$.  The true density (approximated numerically) for this parameter configuration is shown in Figure \ref{fig:gandk_density}.  For each sample size, we generate 100 independent datasets.  For BSL we use $m = 50$ and for KDE we use $m = 100$.  For the summary statistic based approaches, we use a 3 component Gaussian mixture as the auxiliary model\footnote{For datasets where various numerical issues arise, we use 2 components}.  We find MMD to be too slow for the $n = 1000$ datasets.

The results for $n = 100$ are shown in Tables \ref{tab:gandk-100-a}-\ref{tab:gandk-100-k} for the four parameters.  In the tables we show, based on the 100 simulated datasets, the estimated bias of the posterior mean, bias of the posterior median, average of the posterior standard deviation, and the coverage rates for nominal rates of 80\%, 90\% and 95\%. If there is a method that clearly performs best for a particular parameter based on a combination of the performance measures, then we bold it in the table.  We also use italics for any methods that perform relatively well.  We note that there is some subjectivity in these decisions.

\begin{table*}
	\caption{Repeated simulation results for parameter $a$ of the g-and-k example based on simulated data of size $n=100$.}
	\centering%
	\begin{tabular}{ccccccc}
		&	bias (mean) & bias (median) & std & 80\% &  90\% & 	95\% \\
		exact & 0.03 & 0.02 & 0.12   & 86 & 92 & 98  \\
		\textbf{CvM} & 0.001 & -0.008 & 0.12  & 84 & 92 & 99  \\
		Wass & -0.07 & -0.08 & 0.14  & 84 & 95 & 99  \\
		MMD & -0.09 & -0.11 & 0.12  & 76  & 88  & 97  \\
		\textit{KDE} & -0.02 & -0.03 & 0.12  & 79 & 89  & 94 \\
		ABC  & -0.08 & -0.1 & 0.15  & 83 & 95 & 97 \\
		\textit{BSL} & -0.04 & -0.05  & 0.13  & 83 & 91 & 96  \\
	\end{tabular}%
	\label{tab:gandk-100-a}%
\end{table*}

\begin{table*}
	\caption{Repeated simulation results for parameter $b$ of the g-and-k example based on simulated data of size $n=100$.}
	\centering%
	\begin{tabular}{ccccccc}
		&	bias (mean) & bias (median) & std & 80\% &  90\% & 	95\% \\
		exact & 0.08 & 0.06 & 0.23   & 85 & 93 & 98  \\
		\textit{CvM} & 0.05 & 0.02 & 0.26  & 90 & 98 & 100  \\
		Wass & 0.07 & 0.04 & 0.28  & 89 & 96 & 100  \\
		\textit{MMD} & 0.04 & 0.02 & 0.28  & 88  & 98  & 99  \\
		KDE & 0.12 & 0.09 & 0.26  & 81 & 91  & 96 \\
		ABC  & 0.05 & 0.02 & 0.30  & 92 & 97 & 100 \\
		BSL & 0.09 & 0.06  & 0.27  & 85 & 94 & 99  \\
	\end{tabular}%
	\label{tab:gandk-100-b}%
\end{table*}

\begin{table*}
	\caption{Repeated simulation results for parameter $g$ of the g-and-k example based on simulated data of size $n=100$.}
	\centering%
	\begin{tabular}{ccccccc}
		&	bias (mean) & bias (median) & std & 80\% &  90\% & 	95\% \\
		exact & 0.13 & 0.04 & 0.51   & 83 & 92 & 97  \\
		\textbf{CvM} & 0.4 & 0.2 & 0.87  & 89 & 97 & 99  \\
		Wass & 2.2  & 1.5 & 2.5  & 80 & 98 & 98  \\
		MMD & 3.1 & 2.8 & 2.6  & 60  & 86  & 98  \\
		\textit{KDE} & 1.0 & 0.62 & 1.4  & 67 & 82  & 90 \\
		ABC  & 2.1 & 1.5 & 2.3  & 82 & 95 & 98 \\
		BSL & 1.7 & 1.1  & 2.1  & 70 & 88 & 94  \\
	\end{tabular}%
	\label{tab:gandk-100-g}%
\end{table*}

\begin{table*}
	\caption{Repeated simulation results for parameter $k$ of the g-and-k example based on simulated data of size $n=100$.}
	\centering%
	\begin{tabular}{ccccccc}
		&	bias (mean) & bias (median) & std & 80\% &  90\% & 	95\% \\
		exact & -0.02 & -0.03 & 0.13  & 88 & 93 & 96  \\
		CvM & 0.04 & 0.02 & 0.22  & 95 & 100 & 100  \\
		Wass & -0.01  & -0.03 & 0.19  & 89 & 96 & 100  \\
		MMD & 0.01 & -0.01 & 0.24  & 96  & 100  & 100  \\
		\textbf{KDE} & -0.04 & -0.05 & 0.15  & 83 & 90  & 94 \\
		ABC  & -0.07 & -0.09 & 0.20  & 89 & 99 & 100 \\
		\textit{BSL} & -0.05 & -0.06  & 0.15  & 82 & 89 & 97  \\
	\end{tabular}%
	\label{tab:gandk-100-k}%
\end{table*}	

Taking the four parameters into account, CvM could be considered the best performing method.  This approach clearly produces the best results for $g$, which is the most difficult parameter to estimate in the g-and-k model.  However, it generally produces overcoverage.  KDE also generally performs relatively well, followed by BSL.  Wass, MMD and ABC perform relatively poorly in this example.  The Wasserstein distance is likely having difficulty handling the heavy tailed nature of the data.  The results for the larger $n = 1000$ sized datasets (Tables \ref{tab:gandk-1000-a}-\ref{tab:gandk-1000-k}) are qualitatively similar, but the difference between the methods is more subtle.  CvM, KDE and BSL perform similarly, with Wass and ABC noticeably performing worse.  Results for estimated posterior correlations are provided in Appendix A of the supplementary material.  Wass and MMD appear to be the least accurate in recovering the exact posterior correlations in general, which is consistent with the marginal results presented here.


\begin{table*}
	\caption{Repeated simulation results for parameter $a$ of the g-and-k example based on simulated data of size $n=1000$.}
	\centering%
	\begin{tabular}{ccccccc}
		&	bias (mean) & bias (median) & std & 80\% &  90\% & 	95\% \\
		exact & 0.002 & 0.001 &  0.035   & 81 & 94 & 95  \\
		\textit{CvM} & 0.001 & -0.0000 & 0.038  & 87 & 94 & 97  \\
		Wass & -0.003 & -0.003 & 0.044  & 92 & 96 & 99  \\
		\textit{KDE} & -0.009 & -0.01 & 0.035  & 87 & 92  & 94 \\
		ABC  & 0.0000 & -0.001 & 0.044  & 89 & 97 & 99 \\
		\textit{BSL} & 0.0004 & -0.0003  & 0.037  & 83 & 92 & 96  \\
	\end{tabular}%
	\label{tab:gandk-1000-a}%
\end{table*}	

\begin{table*}
	\caption{Repeated simulation results for parameter $b$ of the g-and-k example based on simulated data of size $n=1000$.}
	\centering%
	\begin{tabular}{ccccccc}
		&	bias (mean) & bias (median) & std & 80\% &  90\% & 	95\% \\
		exact & 0.007 & 0.005 & 0.072   & 84 & 95 & 99  \\
		\textit{CvM} & 0.006 & 0.003 & 0.078  & 88 & 98 & 99  \\
		Wass & 0.02 & 0.01 & 0.085  & 88 & 98 & 100  \\
		\textit{KDE} & 0.02 & 0.01 & 0.076  & 85 & 96  & 99 \\
		ABC  & 0.01 & 0.01 & 0.091  & 91 & 99 & 100 \\
		\textit{BSL} & 0.01 & 0.007  & 0.077  & 83 & 92 & 97  \\
	\end{tabular}%
	\label{tab:gandk-1000-b}%
\end{table*}	

\begin{table*}
	\caption{Repeated simulation results for parameter $g$ of the g-and-k example based on simulated data of size $n=1000$.}
	\centering%
	\begin{tabular}{ccccccc}
		&	bias (mean) & bias (median) & std & 80\% &  90\% & 	95\% \\
		exact & 0.004 & 0.0008 & 0.10  & 78 & 86 & 92  \\
		\textit{CvM} & 0.02 & 0.01 & 0.14  & 91 & 97 & 99  \\
		Wass & 0.07 & 0.04 & 0.24   & 97 & 97 & 100  \\
		KDE & 0.10 & 0.08 & 0.12  & 65 & 75  & 83 \\
		ABC  & 0.02 & 0.009 & 0.17  & 81 & 97 & 99 \\
		\textit{BSL} & 0.02 & 0.01  & 0.14  & 75 & 87 & 95  \\
	\end{tabular}%
	\label{tab:gandk-1000-g}%
\end{table*}

\begin{table*}
	\caption{Repeated simulation results for parameter $k$ of the g-and-k example based on simulated data of size $n=1000$.}
	\centering%
	\begin{tabular}{ccccccc}
		&	bias (mean) & bias (median) & std & 80\% &  90\% & 	95\% \\
		exact & -0.002 & -0.003 & 0.041  & 81 & 92 & 100  \\
		CvM & 0.005 & 0.003 & 0.064  & 87 & 100 & 100  \\
		Wass & -0.01 & -0.01 & 0.059  & 97 & 97 & 99  \\
		\textit{KDE} & -0.01 & -0.01 & 0.044  & 84 & 93  & 98 \\
		ABC  & -0.0142 & -0.02 & 0.056  & 92 & 99 & 100 \\
		\textit{BSL} & -0.007 & -0.009  & 0.046  & 80 & 92 & 97  \\
	\end{tabular}%
	\label{tab:gandk-1000-k}%
\end{table*}


\subsection{M/G/1 Example}

The M/G/1 queueing model is a stochastic single-server queue model with Poisson arrivals and a general service time distribution. Here we assume that service times are $\mathcal{U}(\theta_1,\theta_2)$, as this has been a popular choice in other likelihood-free literature (see e.g.\ \citet{an2020robust,Blum2009}). The time between arrivals is $\text{Exp}(\theta_3)$ distributed.  We take the observed data $\y$ to be the inter-departure times of 51 customers, resulting in 50 observations. The observed data is generated with true parameter  $(\theta_1,\theta_2,\theta_3)^\top = (1,5,0.2)^\top$. The prior is $\mathcal{U}(0,\min(y_1,y_2,\ldots,y_n))\times \mathcal{U}(0,10+\min(y_1,y_2,\ldots,y_n)) \times \mathcal{U}(0,0.5)$ on $(\theta_1,\theta_2,\theta_3)$. \cite{shestopaloff2014bayesian} develop a data augmentation MCMC method to sample from the true posterior that we compare the approximate methods with.

\begin{figure}[h]
	\centering
	\includegraphics[scale=0.35]{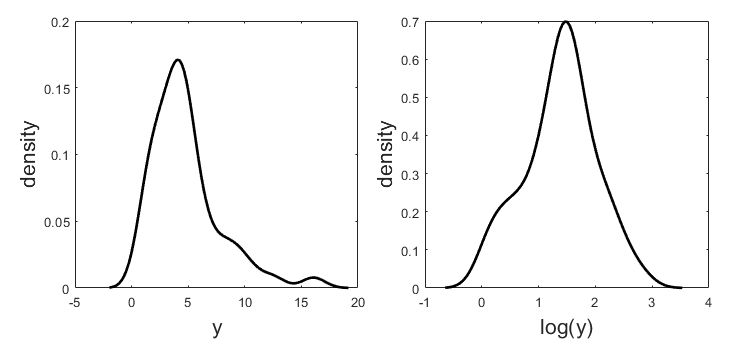}
	\caption{Visualisation of the data simulated from the M/G/1 model. Shown on the left is a kernel density estimate of the underlying inter-departure distribution based on 50 observations.  The right shows the same figure but for the log inter-departure times.}
	\label{fig:mg1_data}
\end{figure}

We compare methods using 100 independent datasets generated from the true M/G/1 model.  A visualisation of one of these datasets is shown in Figure \ref{fig:mg1_data}.  It can be seen that the data shows some positive skewness.  Thus we also consider the impact on the likelihood-free approaches by applying a log transformation to the data, which results in a more symmetric distribution (also shown in Figure \ref{fig:mg1_data}).  The results based on the log transformation data include log in parentheses after the acronym of the method.  The CvM distance is theoretically unaffected by one-to-one transformations of the data, so we expect similar results for both datasets for that approach.  However, the other approaches may be impacted by the transformation.  For the auxiliary model for ABC and BSL we again use a 3 component Gaussian mixture\footnote{again, 2 components when numerical issues are encountered}.  BSL uses $m = 50$ and KDE uses $m = 100$.

The results for the three parameters are shown in Tables \ref{tab:mg1-t1}-\ref{tab:mg1-t3}.  Overall the best performing method for this example is KDE.  There is little to no autocorrelation in the data, justifying the independence assumption of KDE.  Even though the data is skewed, the underlying distribution of the inter-departure time does not have thick tails, and the log transformation helps to remove a large degree of the skewness.  BSL also performs relatively well on this example.  BSL performs substantially better than ABC with the same summary statistics.  This is consistent with the empirical results of \citet{price2018bayesian}, which shows that BSL can outperform ABC when the summary statistic distribution is regular enough.

Interestingly, despite being one of the best performing methods in the g-and-k example, CvM is one of the worst performing in this example.  The results are very similar when the data is log transformed, as expected.  For the other methods, there is generally an improvement in results when log transforming the data, except for MMD, ABC and KDE where the results are worse for $\theta_2$.  The best performing full distance ABC method is Wass (log), with the log transformation being critical to obtain good results for $\theta_1$.  

Unlike in the g-and-k example, the Wasserstein approach significantly outperforms the CvM approach in the M/G/1 example for the parameters $\theta_1$ and $\theta_2$ (the two give largely similar results for $\theta_3$, with the Wasserstein having a slight edge). We hypothesize that this poor performance is due to the relatively small sample size ($n$ = 50 observations); the fact that the parameter $\theta_1$ controls the lower tail of the observed data; and the fact that, due to the large mean of the exponentially distributed interarrival times, the estimated CDF can be quite noisy at large values in the sample.  Results for estimated posterior correlations are provided in Appendix B of the supplementary material.  Most methods do a reasonable job in recovering the exact posterior correlations, except that CvM does not accurately estimate the correlation between $\theta_2$ and $\theta_3$.


\begin{table*}
	\caption{Repeated simulation results for parameter $\theta_1$ of the mg1 example based on simulated data.}
	\centering%
	\begin{tabular}{ccccccc}
		&	bias (mean) & bias (median) & std & 80\% &  90\% & 	95\% \\
		exact & -0.025 & 0.023 & 0.15  & 82 & 88 & 92  \\
		CvM & -0.24  & -0.18  & 0.26  &  65 & 81 & 89 \\
		CvM (log) & -0.24  & -0.18  & 0.26 & 65  &  82 & 89  \\
		Wass & -0.24  &  -0.19 & 0.26  & 65  & 82  & 90   \\
		Wass (log) & -0.10  &  -0.05 & 0.18  & 78 & 87  &  92 \\
		MMD & -0.24  & -0.20  & 0.25  &  63 & 79  & 90  \\
		MMD (log) & -0.14  & -0.10  & 0.20  & 75  & 82  & 90 \\
		ABC  & -0.16  & -0.12  & 0.22  & 69 & 84 & 90 \\
		ABC (log) & -0.083  & -0.041  & 0.17  & 80 & 86 & 92 \\
		KDE & -0.096 & -0.047 & 0.18 & 79  & 89   & 93  \\
		\textbf{KDE (log)} & -0.052 &  -0.0099 & 0.15 & 82  & 88  & 93   \\
		BSL &  -0.14 &  -0.089  & 0.22  &  76 & 84  & 90   \\
		\textit{BSL (log)} & -0.062  &  -0.010 & 0.17 & 79  & 89  & 92   \\
	\end{tabular}%
	\label{tab:mg1-t1}%
\end{table*}

\begin{table*}
	\caption{Repeated simulation results for parameter $\theta_2$ of the mg1 example based on simulated data.}
	\centering%
	\begin{tabular}{ccccccc}
		&	bias (mean) & bias (median) & std & 80\% &  90\% & 	95\% \\
		exact & -0.030 & -0.054 & 0.49 & 80 & 92 & 96  \\
		CvM & 0.24   & 0.15   & 0.83  & 91   & 97   &  99  \\
		CvM (log) & 0.22 & 0.15  & 0.83  &  90 & 98  & 99 \\
		Wass & 0.18 & 0.15 & 0.90 & 92 &  98 &  100 \\
		Wass (log) & 0.20  & 0.14 & 0.81 & 86 & 98  & 100  \\
		MMD & 0.16 & 0.14 & 0.80 & 84 & 97   & 99 \\
		MMD (log) & 0.23  &  0.18 & 0.89 & 77 & 97  & 100 \\
		ABC & 0.23 & 0.25 & 0.88 & 86 & 92 & 98  \\
		ABC (log) & 0.37 & 0.34 & 0.93 & 80 & 93 & 96   \\
		\textbf{KDE} &  0.0038 & -0.021  &  0.58 & 83 & 94  & 99  \\
		KDE (log) & 0.11  & 0.069  & 0.65 & 82 &  94  & 98   \\
		BSL  & 0.087   &  0.052 & 0.72  & 77  & 88  & 95   \\
		BSL (log) & 0.020 & -0.027 & 0.69 & 82  & 90  & 93   \\
	\end{tabular}%
	\label{tab:mg1-t2}%
\end{table*}

\begin{table*}
	\caption{Repeated simulation results for parameter $\theta_3$ of the mg1 example based on simulated data.}
	\centering%
	\begin{tabular}{ccccccc}
		&	bias (mean) & bias (median) & std & 80\% &  90\% & 	95\% \\
		exact & 0.0077 & 0.0068 & 0.028 & 75 & 84  & 93  \\
		CvM & 0.014   & 0.010   &  0.038  & 80  & 87  &  93  \\
		CvM (log) & 0.014  & 0.010 & 0.038   & 79  & 87  & 94  \\
		Wass & 0.012  & 0.010 & 0.031 & 77 & 87 &  93 \\
		Wass (log) &  0.012 & 0.010 & 0.033 & 80 & 88  & 94 \\
		MMD & 0.015  & 0.012  & 0.035  & 77 &  88 & 95 \\
		MMD (log) & 0.013  & 0.011  & 0.034 & 79 & 87  & 95  \\
		ABC & 0.049  & 0.046  &  0.046 & 67 & 75 & 81  \\
		ABC (log) &  0.036  & 0.032 & 0.047  & 73 & 81  & 86   \\
		\textit{KDE} & 0.011 & 0.010 & 0.024 & 69 & 81  & 85  \\
		\textit{KDE (log)} & 0.0095   & 0.0084 & 0.025 & 72 & 80 & 84  \\
		\textit{BSL}  &  0.013 & 0.013 &  0.025 & 60  & 73  & 81   \\
		\textit{BSL (log)} & 0.0098  & 0.0089 & 0.028 &  67 & 77  & 88   \\
	\end{tabular}%
	\label{tab:mg1-t3}%
\end{table*}


\subsection{Stereological Extremes Example}

Here we consider an example in stereological extremes, originally explored in the likelihood-free setting by \citet{bortot+cs07}.  During the process of steel production, the occurrence of microscopic particles, called inclusions, is a critical measure of the quality of steel. It is desirable that the inclusions are kept under a certain threshold, since steel fatigue is believed to start from the largest inclusion within the block.  \citet{bortot+cs07} develop a new model for inclusions.  The stochastic model generates a random number of inclusions, and for each inclusion, the largest principal diameter of an ellipsoidal model of the inclusion in the $2$-dimensional cross-section.  We refer the reader to \citet{bortot+cs07} for more details, and \citet{anderson+c02} for an earlier mathematical modelling approach.

\begin{figure}[h]
	\centering
	\subfigure[simulated data]{\includegraphics[scale=0.25]{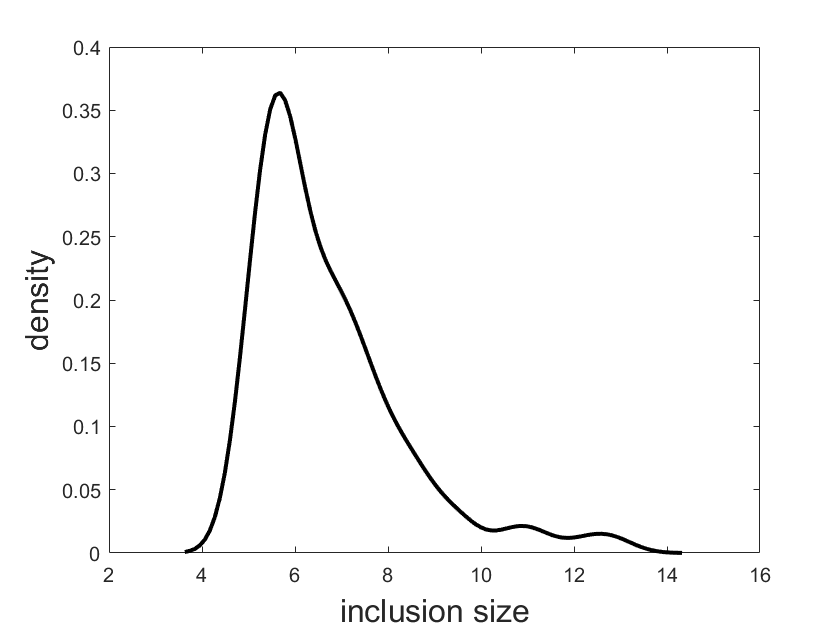}\label{figsub:data_toads_simulated}}
	\subfigure[real data]{\includegraphics[scale=0.25]{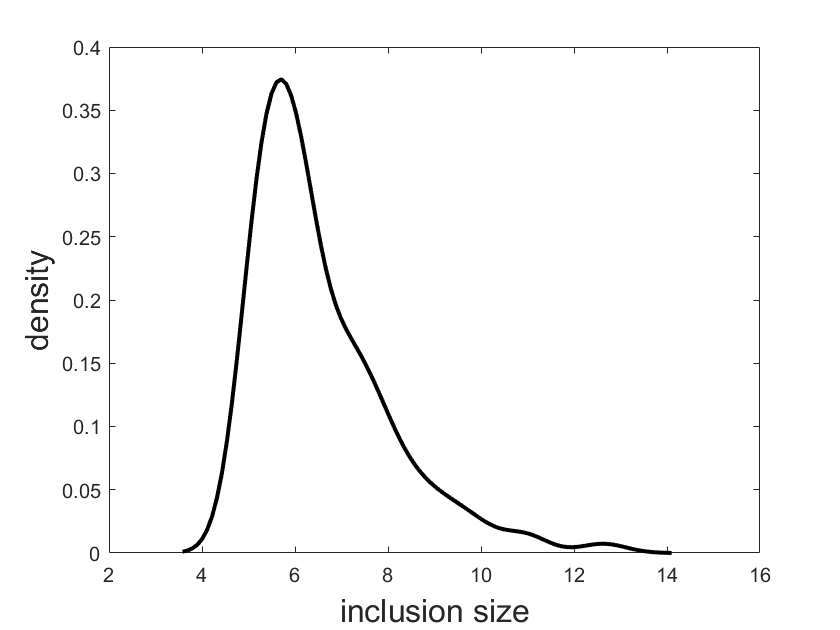}\label{figsub:data_toads_real}}
	\caption{Kernel density estimates of inclusion sizes for the simulated (left) and real (right) data.}
	\label{fig:data_stereo}
\end{figure}

The model contains three parameters, $\theta = (\lambda,\sigma,\xi)$.  Here $\lambda$ is the rate parameter of a homogenous Poisson process describing the locations of the inclusions, and $(\sigma,\xi)$ are the (scale, shape) parameters of a generalised Pareto distribution related to the size of the inclusions.  For more details on the model, see Appendix C of the supplementary material.  The prior distribution is $\mathcal{U}(30,200) \times \mathcal{U}(0,15) \times \mathcal{U}(-3,3)$. If we denote the vector of observed inclusions by $S$, then the observed data is $\y = (S, |S|)$ where $|S|$ represents the number of inclusions.  Here we consider two datasets, the first simulated from the model with true parameter $(100, 2, -0.1)$ and the second being a real dataset as analysed in \citet{bortot+cs07}.  A visualisation of these datasets is shown in Figure \ref{fig:data_stereo}.   The number of inclusions in the simulated and real data is 138 and 112 respectively.

For this application, there has been several sets of summary statistics developed. \citet{Fan2013} consider the number of inclusions, as well as the log of the difference of 112 equally space quantiles, creating 112 statistics in total (111 from the log quantile differences, and the other from the number of inclusions).  \citet{Fan2013} also consider dimension-reduced summary statistics based on the semi-automatic approach of \citet{fearnhead2012constructing}.  We find that similar results can be obtained using summary statistics from our indirect inference approach.  Using a 3 component Gaussian mixture as the auxiliary model produces 9 summary statistics (incorporating the number of inclusions).  We use BSL and ABC with this summary statistic, and for BSL we use $m = 100$ simulated datasets for estimating the synthetic likelihood. We also consider the four summary statistics used in \citet{an2020robust}, which are similar to the original summary statistics in \citet{bortot+cs07}.  These are the number of inclusions, $\log(\min(S))$, $\log(\text{mean}(S))$ and $\log(\max(S))$.  As there are only 4 statistics, we consider only ABC and not BSL.


For the full data distance based ABC approaches, we combine 2 distance functions into a single distance function via a weighted average of individual distances (one for the number of inclusions and one for the inclusion sizes).  For the count of the number of inclusions, we simply use the L1-norm for the distance.  We set the weight for each distance as the inverse standard deviation of the distance estimated from simulations at the true parameter value $(100, 2, -0.1)$.  If the distribution of the distance has a heavy tail, we use a robust estimate of the standard deviation via 1.4826 times the median absolute deviation.

The KDE method is awkward to apply in this example, since the dataset size is random, and thus it is necessary to include not only the inclusion size data but also the number of inclusions.  Here we use $m$ simulated datasets to estimate the density of the number of inclusions.  As each model simulation often generates more than one inclusion, we concatenate all the simulated inclusion sizes together for estimating the density of the inclusion size.  We then treat the number of inclusions and inclusion sizes as independent when estimating the likelihood via KDE.  As with BSL, we use $m = 100$ for KDE.  The MCMC acceptance rate for KDE is higher than BSL with this value of $m$.  However, there is a fair amount of overhead for computing the kernel density estimate for KDE with $m = 100$, so we do not consider larger values of $m$.

\begin{figure*}
	\centering
	\subfigure[full data distances results]{\includegraphics[scale=0.5]{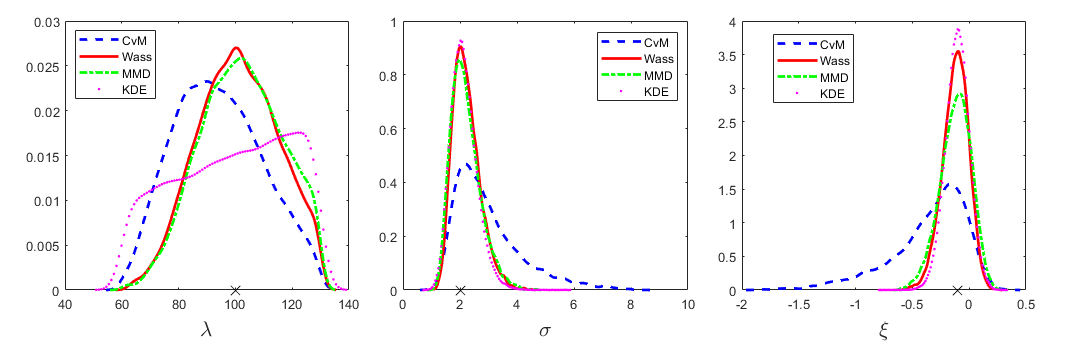}\label{figsub:results_stereo_simulated_dist}}
	\subfigure[summary statistics results including Wass from (a)]{\includegraphics[scale=0.5]{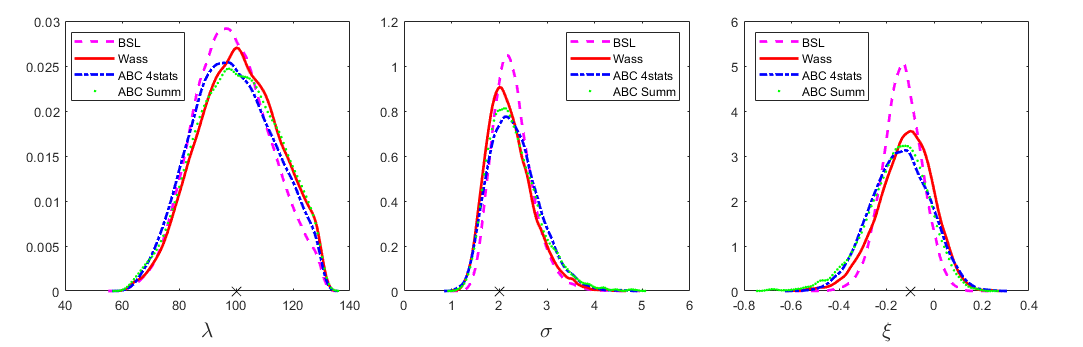}\label{figsub:results_stereo_simulated_summ}}
	\caption{Comparison of estimates of the univariate ABC posterior distributions for the stereological extremes example based on simulated data. Shown are (a) comparisons with distance functions involving the full data and (b) comparisons with summary statistic based approaches.}
	\label{fig:stereo_simulated}
\end{figure*}

\begin{figure*}
	\centering
	\subfigure[full data distances results]{\includegraphics[scale=0.55]{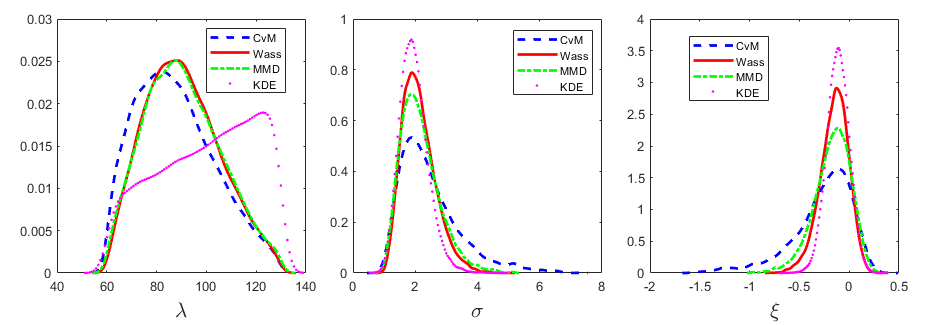}\label{figsub:results_stereo_real_dist}}
	\subfigure[summary statistics results including Wass from (a)]{\includegraphics[scale=0.55]{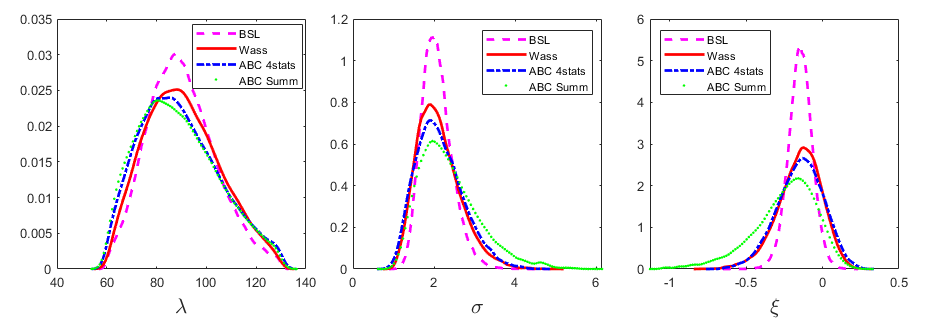}\label{figsub:results_stereo_real_summ}}
	\caption{Comparison of estimates of the univariate ABC posterior distributions for the stereological extremes example based on real data. Shown are (a) comparisons with distance functions involving the full data and (b) comparisons with summary statistic based approaches.}
	\label{fig:stereo_real}
\end{figure*}

The results are shown for the simulated and real datasets in Figures \ref{fig:stereo_simulated} and \ref{fig:stereo_real}, respectively.  In the top row of each figure we compare the full data approaches.  Then, in the second row, we compare the best performing full data approach with the summary statistic approaches.  The results are qualitatively similar for the simulated and real datasets.  For the full data distance approaches, the top performing methods are Wass and MMD.  KDE performs well for $\sigma$ and $\xi$, but not for $\lambda$.  CvM produces the least precise posteriors in general. The poor performance of the CvM is not particularly surprising given the results of the M/G/1 example. In particular, the parameters $(\sigma,\xi)$ control the tail shape of the distribution, and the sample size is relatively small. As we have already discussed, the CvM distance can be quite noisy in these circumstances, and thus the resulting posteriors can be inaccurate. As the Wasserstein distance is more convenient compute than the MMD, we take the Wass method forward to compare with the summary statistic based approaches.  Wass performs similarly to ABC with summary statistics (both choices of the summary statistics).  However, BSL appears to produce slightly more precise posteriors, particularly for $\xi$.


\subsection{Toad Example}

The next example we consider is the individual-based movement model of Fowler's Toads (\textit{Anaxyrus fowleri}) developed by \citet{marchand2017stochastic}. The model has since been considered as a test example in likelihood-free literature, in particular for synthetic likelihood methods (see \citealt{an2020robust,frazier2019robust,priddle2019efficient}).  We consider the ``random return" model of   \citet{marchand2017stochastic}. We provide only a brief overview of the model herein, and refer the reader to \cite{marchand2017stochastic} for more details.
For a particular toad, we draw an overnight displacement from the Levy alpha-stable distribution $S(\alpha, \xi)$, where $0\leq\alpha\leq 2$ and $\xi>0$. At the end of the night, toads return to their previous refuge site with probability $p_0$, or take refuge at their current overnight displacement. In the event of a return on day $i$, the refuge site is chosen with probability proportional to the number of times the toad has previously each refuge site. The raw data consists of the refuge locations of $n_t = 66$ toads over $n_d = 63$ days.  We consider both real and simulated data.  The simulated data is generated using $\theta = (\alpha,\xi,p_0)^\top = (1.7,35,0.6)^\top$, which seems to be also favourable for the real data. 

\begin{figure*}
	\centering
	\subfigure[simulated data]{\includegraphics[scale=0.6]{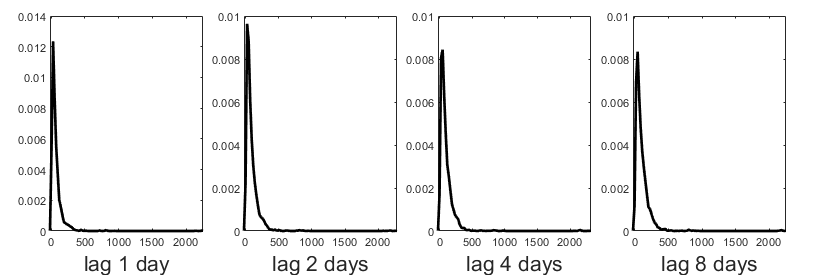}\label{figsub:data_toads_simulated}}
	\subfigure[real data]{\includegraphics[scale=0.6]{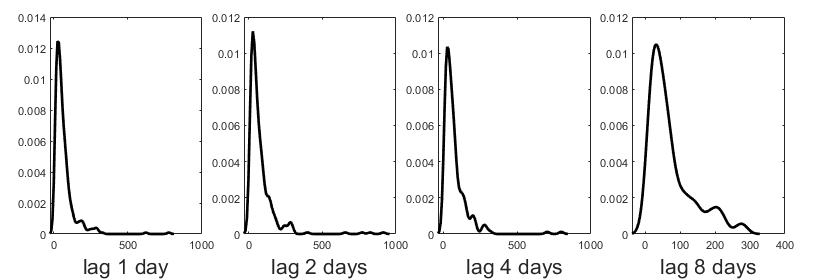}\label{figsub:data_toads_real}}
	\caption{Kernel density estimates of the non-returns distributions for lags 1, 2, 4 and 8 days.  The top row is for simulated data and the bottom row is for real data.}
	\label{fig:data_toad}
\end{figure*}

The raw data consist of GPS location data for $n_t$ toads for $n_d$ days, i.e.\ the observation matrix $\vect{Y}$ is of dimension $n_d \times n_t$. Here $n_t=66$, $n_d=63$.  Unlike the previous examples, we compute an initial set of summary statistics as in \citet{marchand2017stochastic}.  Specifically, $\vect{Y}$ is summarised down to four sets comprising the relative moving distances for time lags of $1,2,4,8$ days. For instance, $\vect{y}_1$ consists of the displacement information of lag $1$ day, $\vect{y}_1 = \{|\Delta y| = |\vect{Y}_{i,j}-\vect{Y}_{i+1,j}| ; 1 \leq i \leq n_d-1, 1 \leq j \leq n_t \}$.  For each lag, we split the displacement vector into two sets.  The first set holds displacements less than 10m, and these are taken as returns, and we simply record the number of returns.  The second set holds the vector of displacements that are greater than 10m (non-returns).  Combining these two aspects (the number of returns and the vector of non-return displacements) for the four lags produces the data $\y$ for analysis.  A visualisation of the non-returns data is given in Figure \ref{fig:data_toad}.  It can be seen that the non-returns data has a heavy right tail.  The number of returns is in the order of 1000 for the simulated data and 100 for the real data.  This is because there are many missing distances in the real data.

For the full distance based ABC approaches, there is no further dimension reduction.  We find that standard BSL is not suitable when the summary statistics are formed from a Gaussian mixture model due to lack of normality.  Instead, we use the statistics from \citet{an2020robust} as BSL appears to work well with them.   For the non-returns, we compute the log of the differences of the $0,0.1,\dots,1$ quantiles and the median for each lag. Combined with the statistics for the returns, there are 48 summary statistics in total.  For BSL we use $m = 500$.  For ABC with summary statistics, we use a weighted euclidean distance, where the weights are the inverse of the standard deviations of the summary statistics estimated from pilot simulations at $(1.7,35,0.6)$.  Appendix D of the supplementary material provides more detail on the model and summary statistics.

For the full data distance based ABC approaches, we combine 8 distance functions into a single distance function via a weighted average of individual distances (returns and non-returns for the four lags).  For the count of the number of returns, we simply use the L1-norm for the distance.  Given the heavy tail nature of the data, we also consider the log of the non-returns for the Wass and MMD.  Given the relatively large number of non-returns in the simulated data we find MMD to be too slow.   Also, the KDE method is awkward to apply for the same reason as the stereological extremes example.  With a moderate value of $m$ needed to estimate the density for the number of returns, a huge number of non-returns is generated and the kernel density estimate is expensive to compute.  Thus we do not consider the KDE method here.

\begin{figure*}
	\centering
	\subfigure[full data distances results]{\includegraphics[scale=0.5]{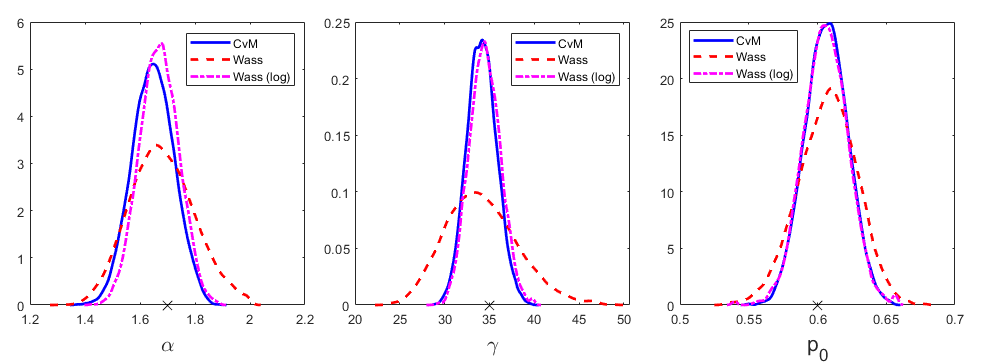}\label{figsub:results_toad_simulated_dist}}
	\subfigure[summary statistics results including CvM from (a)]{\includegraphics[scale=0.5]{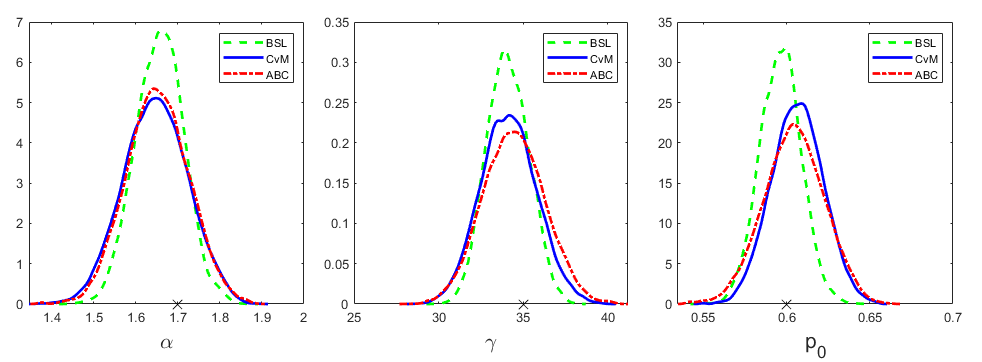}\label{figsub:results_toad_simulated_summ}}
	\caption{Comparison of estimates of the univariate ABC posterior distributions for the toad example based on simulated data. Shown are (a) comparisons with distance functions involving the full data and (b) comparisons with summary statistic based approaches.}
	\label{fig:toad_simulated}
\end{figure*}
\begin{figure*}
	\centering
	\subfigure[full data distances results]{\includegraphics[scale=0.5]{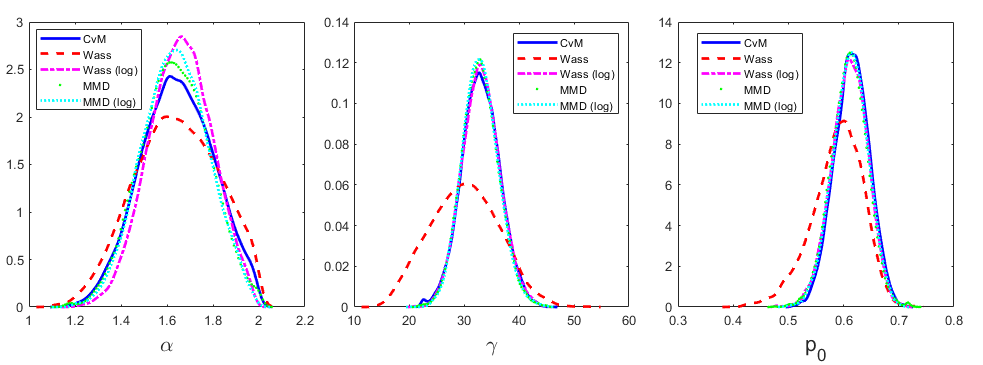}\label{figsub:results_toad_real_dist}}
	\subfigure[summary statistics results including CvM from (a)]{\includegraphics[scale=0.5]{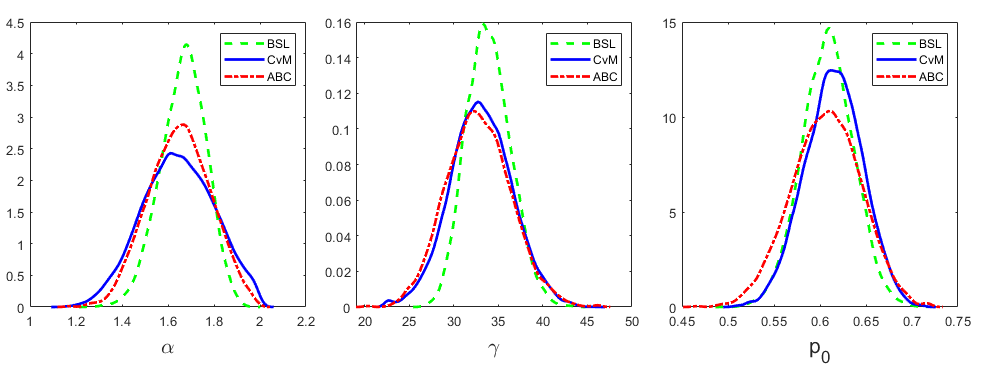}\label{figsub:results_toad_real_summ}}
	\caption{Comparison of estimates of the univariate ABC posterior distributions for the toad example based on real data. Shown are (a) comparisons with distance functions involving the full data and (b) comparisons with summary statistic based approaches.}
	\label{fig:toad_real}
\end{figure*}

The estimated posterior marginals for the simulated data and real data are shown in Figures \ref{fig:toad_simulated} and \ref{fig:toad_real}, respectively.  The top row in each figure compares the full distance based approaches, and then the bottom row compares the best performing full distance approach with the summary statistic based methods.  Appendix D of the supplementary material shows the estimated bivariate posteriors for all the methods.

For both the simulated and real data it is evident that the full distance based approaches perform similarly, except for Wass, which performs particularly poorly for $\gamma$.  As with the M/G/1 example, it is interesting that performing a log transform of the data significantly improves the performance of the Wasserstein distance. In contrast to the Wasserstein distance,  the heavy tailed nature of the data does not affect the results based on the CvM distance. This finding is unsurprising since the CvM distance is generally robust to heavy tailed data. Given that the CvM performs relatively well from a statistical and computational perspective, and it does not require choosing a data transform, we take this method forward to compare with the summary statistic based approaches.  ABC with summary statistics produces similar results to CvM.  BSL generally produces more precise inferences compared to all other methods.

\subsection{Toggle Switch Example}

\begin{figure}[h]
	\centering
	\includegraphics[scale=0.2]{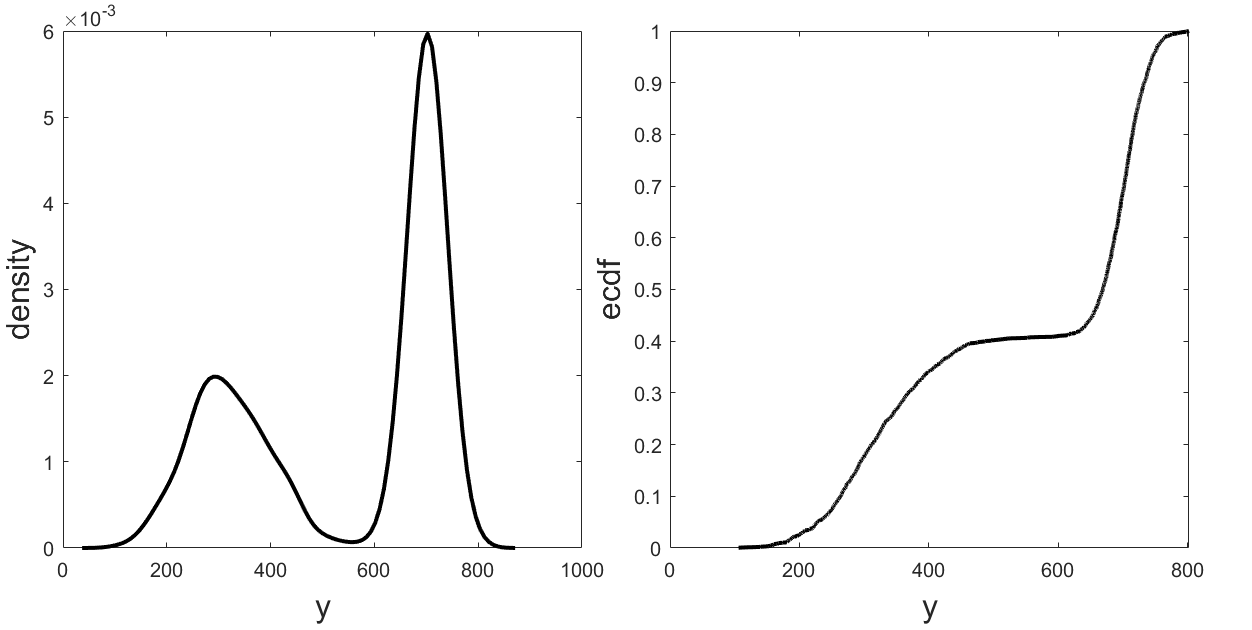}
	\caption{Visualisation of the data simulated from the toggle switch model. Shown on the left is a kernel density estimate based on data from 2000 cells, and the right plot shows the corresponding empirical cumulative distribution function.}
	\label{fig:toggle_data}
\end{figure}

We also consider a toggle switch model describing gene expressions that can produce multi-modal data.  Here we briefly describe  the model, and refer to \citet{Bonassi2011} and \citet{Gardner2000} for more details.  The example has been considered in a likelihood-free context by, for example, \citet{Bonassi2015} and \citet{Vo2019}.  Let $u_{c,t}$ and $v_{c,t}$ be the expressions of genes $u$ and $v$ for cell $c$ at time $t$.  We assume there are 2000 independent cells, $c = 1,\ldots,2000$.  Given an initial state $(u_{c,0}, v_{c,0})$ and a discrete time step $h$,  $u_{c,t}$ and $v_{c,t}$ evolve according to:
\begin{align*}
u_{c,t+h} &= u_{c,t} + h\alpha_u/(1 + v_{c,t}^{\beta_u}) - h(1 + 0.03u_{c,t}) + \\
& \quad 0.5h \xi_{c,u,t}, \\
v_{c,t+h} &= v_{c,t} + h\alpha_v/(1 + c_{c,t}^{\beta_v}) - h(1 + 0.03v_{c,t}) + \\
& \quad 0.5h \xi_{c,v,t},
\end{align*}
where $\xi_{c,u,t}$ and $\xi_{c,v,t}$ are independent standard normal random variates that represent the intrinsic noise within cell $c$.  The observed data consists of noisy measurements of $\{u_{c,T}\}_{c=1}^{2000}$ for some steady state time $T$. The observation for each cell is modelled as
\begin{align*}
y_c &= u_{c,T} + \mu + \mu \sigma \eta_c/u_{c,T}^{\gamma}, \quad \mbox{for } c = 1,\ldots,2000,
\end{align*}
where the errors $\eta_c$ have a standard normal distribution. Therefore the data consist of independent observations, $\y = (y_1,y_2,\ldots,y_{2000})$.  The unknown parameter is $\theta = (\mu, \sigma, \gamma, \alpha_u, \beta_u, \alpha_v, \beta_v)$, and we set $h=1$ and $T = 300$ as in \citet{Bonassi2011}.  We use the same priors as in \citet{Bonassi2011}, which are independent and uniformly distributed with lower and upper bounds of $(250, 0.05, 0.05, 0, 0, 0, 0)$ and $(400, 0.5, 0.35, 50, 7, 50, 7)$, respectively.  The observed data is simulated from the model with true parameter $\theta = (320, 0.25, 0.15, 25, 4, 15, 4)$.  A visualisation of the data is shown in Figure \ref{fig:toggle_data}.  Given the relatively large size of the data, we find MMD and KDE substantially more computationally expensive than the other approaches, so we do not consider them in this example.  For ABC and BSL we use an auxiliary 3 component Gaussian mixture model.

\begin{figure*}
	\centering
	\includegraphics[scale=0.45]{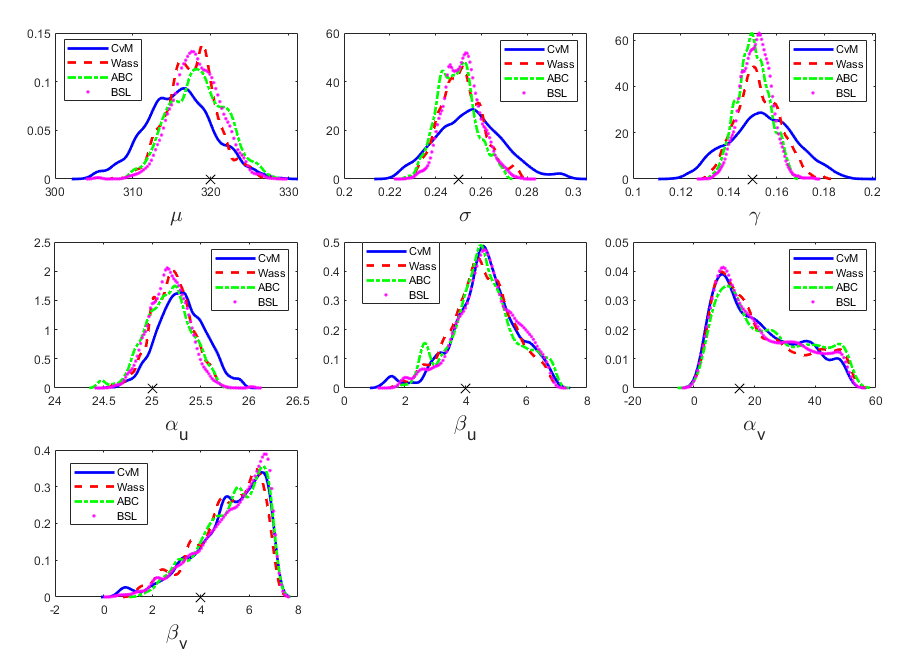}
	\caption{Comparison of estimates of the univariate likelihood-free posterior distributions for the toggle switch example based on simulated data. }
	\label{fig:results_toggle}
\end{figure*}   

Estimates of the univariate posterior distributions are shown in Figure \ref{fig:results_toggle}.  It can be seen that all methods perform similarly, except that the CvM produces more diffuse posteriors for $\mu$, $\sigma$ and $\gamma$.  Results for other simulated datasets generated from the prior predictive distribution are shown in the Appendix E of the supplementary material, and present a wide variety of features. On these additional datasets, the only method to perform consistently well is BSL. CvM generally produces the worst performance, but ABC and Wass also produce poor results in some instances.

We hypothesise that the relatively poor performance of CvM, for $\mu$, $\sigma$ and $\gamma$, is due to the existence of flat regions in the ECDF for data generated from the toggle model; for example, in Figure \ref{fig:toggle_data} the ECDF is nearly flat around a probability of 0.4. Since $\mu$, $\sigma$ and $\gamma$ directly influence the observed data, $y_c$, this flat region suggests that there are many different values of these parameters for which the ECDF is similar, which would likely result in diffuse posteriors when the chosen distance is CvM. Further evidence for this hypothesis is shown in the Appendix for datasets with similar features.  The results in the Appendix also show that CvM can produce relatively poor results for $\alpha_u$ and $\beta_v$.  We find that this is particularly the case for datasets where there exist a small number of observations far from the bulk of the data.  From some model simulations we find that $\alpha_u$ and $\beta_v$ can have a strong influence on the presence of these `outliers'.  The CvM distance places little emphasis on these outliers and thus can accept values of $\alpha_u$ and $\beta_v$ that produce simulated data with no outliers, whereas other distances reject these datasets, leading to relatively diffuse approximate posteriors.  It is also worth noting that Wass also produces relatively poor results for $\alpha_u$ for these datasets with outliers.  In contrast to CvM, Wass places too much emphasis on the outliers, again leading to an approximate posterior that is too diffuse.

\section{Discussion} \label{sec:discussion}

In this article we reviewed likelihood-free approaches that avoid data summarisation, predominantly focussing on full data distance functions in the ABC context.  We performed a qualitative and quantitative comparison between these methods.  This should assist practitioners in choosing distance functions that are likely to perform relative well for their specific applications and data types. We also extended the comparison to likelihood-free approaches that resort to data reduction.  We found that at least one of the full data approaches was competitive with or outperforms ABC with summary statistics across most examples, except for some datasets in the toggle switch example.  Another interesting finding is that the performance of the full data approaches can be greatly affected by data transformations.  The CvM distance function is appealing as it is invariant to monotone transformations, it is fast to compute and is more widely applicable than other full data approaches.  However, it did not perform well for the M/G/1 and stereological extremes examples.  It would not be difficult to run ABC with different choices of the full data distance functions on parallel cores.    

Another finding of this research is that full data distances may need to be split, or augmented with additional information, to ensure they can identify all the model parameters. For example, in the invasive toad model, the data on returns and non-returns, as well as their lags, carry specific information about the model parameters, and combining these data within a single distance can result in a loss of information for certain model parameters; hence, in this example we combine eight different full data distances - one distance for each the first four lags of the returns and non-returns series - to conduct posterior inference. Furthermore, there are other cases, such as the stereological extremes example, where is it useful to augment the full data distances with summary statistics (a finding that was first noted in the case of the Wasserstein distance by \citealp{Bernton2019}); in this example the number of inclusions carries important information that cannot be recovered by a distance based solely on the inclusion sizes. We suggest, as a default, using a weighted average when combining multiple distance functions, where the weights for each distance are inversely proportional to the standard deviation of the distance.

We note that BSL performed well across all the examples, but it relies on a Gaussian assumption of the model summary statistic, either the full distribution or the dependence structure. From these examples, it is reasonable to hypothesize that if a great deal of effort is placed on finding informative summary statistics, approaches that using these summaries are likely to outperform the full data approaches in many applications.  However, the promising performance of the full data approaches warrants further research in this direction, especially considering that these methods completely obviate the need to choose summary statistics. 

We also note that the results obtained for ABC with summary statistics could possibly be improved with the use of regression adjustment methods,  see \cite{blum2017regression} for a review of such methods, or methods that first conduct summary statistic selection (\citealp{Prangle2018}). However, such adjustments have not been considered herein to facilitate a more direct comparison between summary-based ABC and ABC based on full data distance approaches.

In some examples we found it useful to combine full data distances with summary statistics that are informative about particular parameters.  We expect this approach to be useful in other applications.  However, how to appropriately weight the different components in the overall ABC discrepancy function is less clear, and the inferences could be sensitive to these weights.  In this paper we adopted a pragmatic approach and set each weight to depend on the variability of its corresponding distance function estimated from simulations at a parameter value within the bulk of the posterior (the true parameter value when it is available).  However, this weighting is not guaranteed to be optimal.  The research in \citet{Prangle2017} and \citet{Harrison2020}, which aim to optimally weight summary statistics within an ABC discrepancy function, could be adapted to the setting of ABC discrepancy functions that combine full data distances and summary statistics. 

For future research we plan to explore these full data distance ABC approaches in the context of likelihood-free model choice.  Problems with performing model choice on the basis of summary statistics have been well documented \citep{Robert2011,Marin2013}.  In this paper we assume that the models are well specified, and are able to capture the characteristics of the observed data. However, an interesting extension of this research would be to perform an extensive comparison of full data and summary statistic based approaches in the setting of model misspecification.  Such analysis would be particularly interesting given that several studies have documented the potential for poor behavior of summary statistic based approaches in misspecified models (see \citealp{frazier2020model} for a discussion in the case of ABC, and \citealp{frazier2019robust} for a discussion in the case of BSL), while the results of \cite{Frazier2020} suggest that full data approaches to likelihood-free inference can deliver inferences that are robust to certain forms of model misspecification.  To further complicate the comparison, a careful choice of summary statistics that ignore features of the data that the model cannot capture can produce inferences robust to misspecification \citep{Lewis2021}.  Given that any ranking between the methods is likely to be example specific, great care would be needed in order to construct a set of examples that is broad enough to cover the most common types of model misspecification encountered in practice. Therefore, we leave this interesting topic for future research.

A limitation of the full data approaches in this paper are the types of observed data they can feasibly handle.  Currently, these methods are predominantly suited to univariate  datasets. {We speculate that the application of these distances to univariate problems is due to both computational and statistical concerns. From a statistical standpoint, it is well-known that the Wasserstein distance has a rate of convergence that depends on the dimension of the data, while the simulated likelihood approach of \cite{Turner2014} encounters a similar curse of dimensionality, since it it based on nonparametric density estimation. While multivariate extensions that can mitigate the impact of data dimension do exist, from a computational standpoint, such procedures can be computationally expensive to implement in higher dimensions, and so to date these procedures have not received much focus in the likelihood-free literature.} In future work, we plan to explore extending the methods to handle higher dimensional data by exploiting recent research on multivariate non-parametric tests (e.g.\ \citealp{Kim2020}). This may increase the class of problems where full data approaches are applicable \citep{Kim2020}.  It would also be interesting to extend and compare methods on temporal and/or spatial data.  This might motivate the use or development of other distance functions.   

\section*{Acknowledgements}
CD gratefully acknowledges support from the Australian Research Council Future Fellowship Award FT210100260.  DT gratefully acknowledges support by the Australian Research Council through
grant DE200101070. The authors acknowledge support by the Australian Centre of Excellence in Mathematics and Statistics.  CD acknowledges support from the QUT Centre for Data Science. Comments from two anonymous referees have led to improvements in this paper.

\bibliographystyle{apalike}
\bibliography{refs}

\newpage

\section*{Appendix A: Additional Results for g-and-k Example}

The estimated posterior correlations between parameters of the g-and-k model produced by various methods for samples sizes $n=100$ and $n=1000$ are shown in Tables \ref{tab:gandk-correlations-100} and \ref{tab:gandk-correlations-1000}, respectively.  These are results averaged over 100 independent datasets.  It can be seen that most methods do a reasonable job of recovering the exact posterior correlations, with Wass and MMD appearing to be the least accurate in general.

\begin{table*}[htp]
	\caption{Average correlation between parameters of the g-and-k example over 100 repeated simulations of size $n=100$.}
	\centering%
	\begin{tabular}{ccccccc}
		&	corr$(a,b)$ & corr$(a,g)$ & corr$(a,k)$ & corr$(b,g)$ & corr$(b,k)$ & corr$(g,k)$ \\
		exact & 0.70 &   -0.39 &   -0.53 &    0.12 &   -0.71 &    0.16 \\
		CvM & 0.64 &   -0.43 &   -0.43 &    0.08 &   -0.61 &    0.26  \\
		Wass & 0.44 &   -0.59 &   -0.28 &   -0.04 &   -0.76 &    0.06   \\
		MMD & 0.51 &   -0.55 &   -0.26 &   -0.06 &   -0.72 &    0.11   \\
		KDE & 0.57 &   -0.48 &   -0.45 &    0.12 &   -0.77 &    0.02  \\
		ABC  &  0.53 &   -0.53 &   -0.35 &   -0.006 &   -0.71 &    0.12
		\\
		BSL & 0.59 &   -0.57 &   -0.46 &   -0.04 &   -0.79 &    0.13   \\
	\end{tabular}%
	\label{tab:gandk-correlations-100}%
\end{table*}

\begin{table*}[htp]
	\caption{Average correlation between parameters of the g-and-k example over 100 repeated simulations of size $n=1000$.}
	\centering%
	\begin{tabular}{ccccccc}
		&	corr$(a,b)$ & corr$(a,g)$ & corr$(a,k)$ & corr$(b,g)$ & corr$(b,k)$ & corr$(g,k)$ \\
		exact & 0.74 &   -0.31 &   -0.60 &    0.18 &   -0.76 &    0.15  \\
		CvM & 0.71 &   -0.39 &   -0.53 &    0.07 &   -0.66 &    0.41  \\
		Wass & 0.50 &  -0.55 &   -0.37 &    0.08 &   -0.78 &    0.11   \\
		KDE & 0.70 &  -0.33 &   -0.58 &    0.23 &   -0.78 &    0.07  \\
		ABC  & 0.65 &   -0.41 &   -0.50 &    0.18 &   -0.77 &    0.10  \\
		BSL & 0.65 &   -0.40 &   -0.51 &    0.19 &   -0.78 &    0.08  \\
	\end{tabular}%
	\label{tab:gandk-correlations-1000}%
\end{table*}

\section*{Appendix B: Additional Results for M/G/1 Example}

The estimated posterior correlations between parameters of the mg1 model produced by various methods is shown in Table \ref{tab:mg1-correlations}. Most methods do a reasonable job of recovering the exact posterior correlations, except for CvM which does not accurately estimate the correlation between $\theta_2$ and $\theta_3$.

\begin{table*}[htp]
	\caption{Average correlation between parameters of mg1 example over 100 repeated simulations.}
	\centering%
	\begin{tabular}{cccc}
		&	corr$(\theta_1,\theta_2)$ & corr$(\theta_1,\theta_3)$ & corr$(\theta_2,\theta_3)$ \\
		exact & -0.02  &  0.006 &   -0.01  \\
		CvM & -0.04 &   -0.008 &    0.19 \\
		CvM (log) & -0.04 &   -0.008 &   0.18  \\
		Wass & -0.08 &  -0.01  &  0.03   \\
		Wass (log) & -0.05 &    0.003 &    0.08 \\
		MMD & -0.07 &   -0.05 &    0.09  \\
		MMD (log) & -0.03  & -0.03 &    0.09 \\
		ABC  & -0.04  &  0.02 &   0.07 \\
		ABC (log) & -0.06  &  0.01 &   0.10 \\
		KDE &  -0.04 &   -0.05 &   -0.01  \\
		KDE (log) & -0.03 &  -0.03 &  -0.0018   \\
		BSL &  -0.04 &  -0.07 &   -0.02  \\
		BSL (log) & -0.04 &   -0.06 &    0.01   \\
	\end{tabular}%
	\label{tab:mg1-correlations}%
\end{table*}	

\section*{Appendix C: Additional Details and Results for Stereological Extremes Example}

Here we follow the description and notation provided in \citet{bortot+cs07}.  \citet{bortot+cs07} consider  analysing the production of clean steal, which can be affected by small particles called inclusions.  It is suggested that the strength of a block of clean steal is affected by the largest inclusion.  Therefore there is interest in analysing the larger inclusions above a threshold $v_0$.  \citet{anderson+c02} assume that inclusions are spherical. Each inclusion diameter $v$ is assumed to be drawn independently from the generalised Pareto distribution with cumulative distribution function:
\begin{align*}
G(v) &= 1 - \left\{1 + \frac{\xi(v-v_0)}{\sigma} \right\}_+^{-1/\xi},
\end{align*}
where $v > v_0$, $\sigma > 0$, $\xi \in \mathbb{R}$ and $a_+ = \max(a,0)$.  The introduction of inclusions with diameter greater than $v_0$ are assumed to come from a Poisson process with rate $\lambda$. The parameter of interest is thus $\theta = (\lambda,\sigma,\xi)$.  The diameters of the inclusions are not directly observed, but the set of 2-dimensional cross-sectional diameters, $S_1,S_2,\ldots,S_n$ are, where the number of inclusions $n$ is a random variable.  The cumulative distribution function associated with each $S_i$ is given by a result due to \citet{Wicksell1925}:
\begin{align}
\mathrm{Pr}(S \leq s | S > v_0) &= 1 - \frac{\int_s^\infty (v^2 - s^2)^{1/2}dG(v)}{\int_s^\infty (v^2 - v_0^2)^{1/2}dG(v)}, \label{eq:Wicksell}
\end{align}  
where $s \geq v_0$.  Associated with each $S_i$ is a latent volume $V_i$.    Based on the spherical inclusion assumption, \citet{anderson+c02} develop a Markov chain Monte Carlo algorithm to infer $\theta$ and the latent $V_i$'s.

Given the unlikely assumption of spherical inclusions, \citet{bortot+cs07} develop a model where the inclusions are ellipsoidal, and where the planar measurement $S_i$ is assumed to be the largest principal diameter of the ellipse generated by the planar section of an inclusion.  Denote $(W_1,W_2,W_3)$ as the principal diameters of a random inclusion, then without loss of generality we re-define $W_3$ as $\max(W_1,W_2,W_3)$.  It is assumed that $W_j = U_j W_3$, where $U_j \sim \mathcal{U}(0,1)$ for $j = 1, 2$.

Unfortunately, there is currently no extension of the result of \eqref{eq:Wicksell} to the ellipsoidal case, which is required for the likelihood-based inference of \citet{anderson+c02} based on the spherical assumption.  However, it is comparatively straightforward to simulate a random set of inclusions under the ellipsoidal model for a given $\theta$.  In the main paper we take $(S,|S|)$ as the summary statistics where $S = (S_1,S_2,\ldots,S_n)$ and $|S|$ is the length of $S$, which is important to consider since, as mentioned before, $n$ is a random variable.

\begin{figure}[H]
	\centering
	\subfigure[CvM]{\includegraphics[scale=0.35]{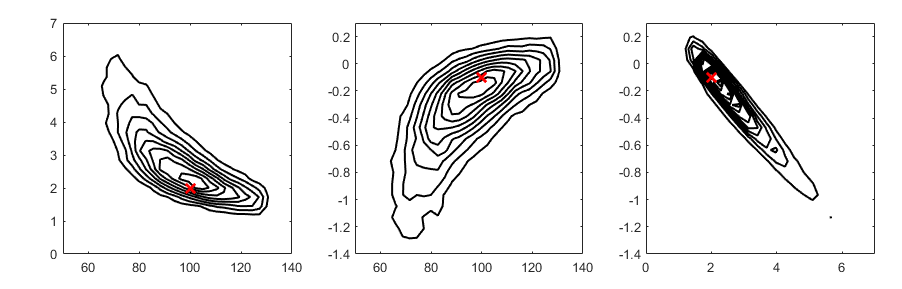}\label{figsub:results_toad_bivariate_simulated_cvm}}
	\subfigure[Wass]{\includegraphics[scale=0.35]{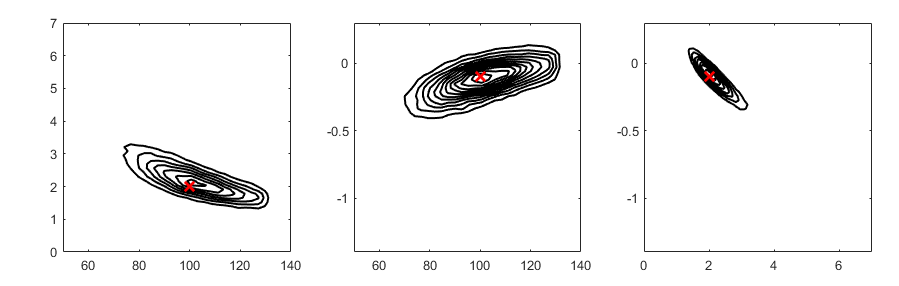}\label{figsub:results_toad_bivariate_simulated_wass}}
	\subfigure[MMD]{\includegraphics[scale=0.35]{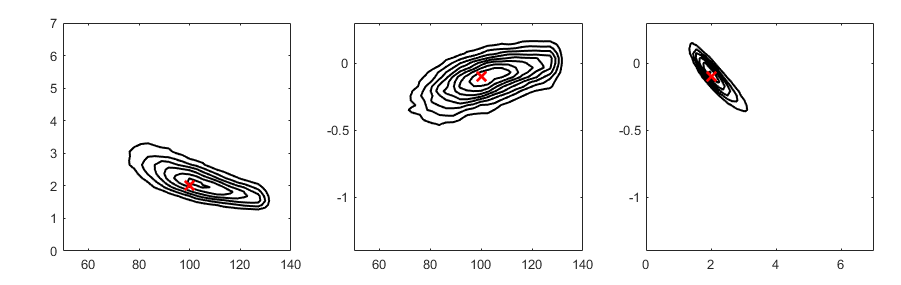}\label{figsub:results_toad_bivariate_simulated_mmd}}
	\subfigure[ABC 4stats]{\includegraphics[scale=0.35]{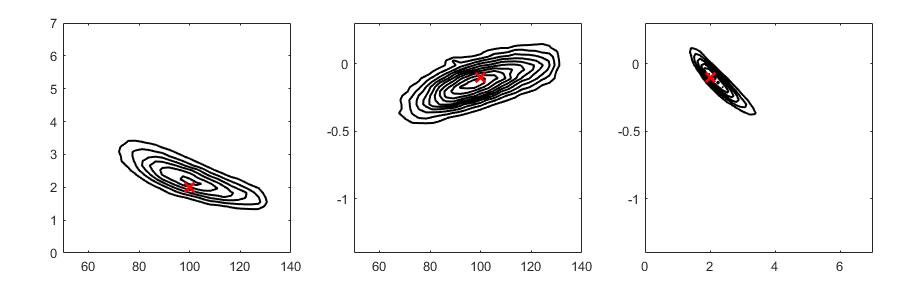}\label{figsub:results_toad_bivariate_simulated_4stats}}
	\subfigure[ABC Summ]{\includegraphics[scale=0.35]{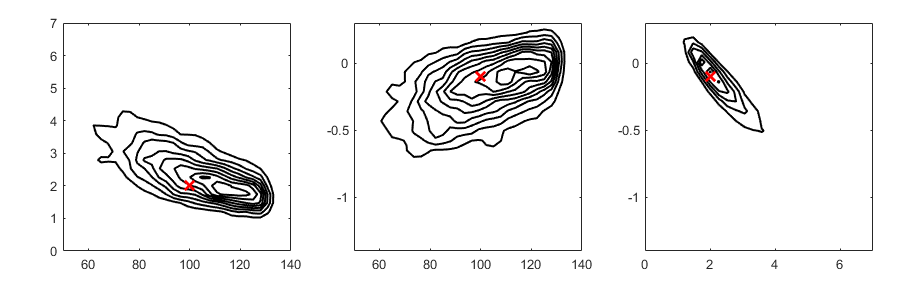}\label{figsub:results_toad_bivariate_simulated_quantiles}}
	\subfigure[KDE]{\includegraphics[scale=0.35]{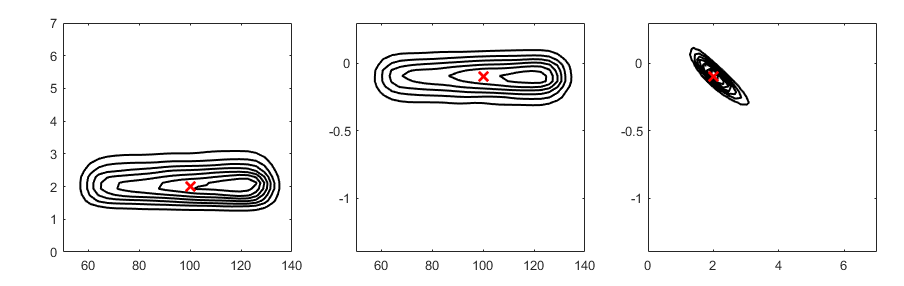}\label{figsub:results_toad_bivariate_simulated_kde}}
	\subfigure[BSL]{\includegraphics[scale=0.35]{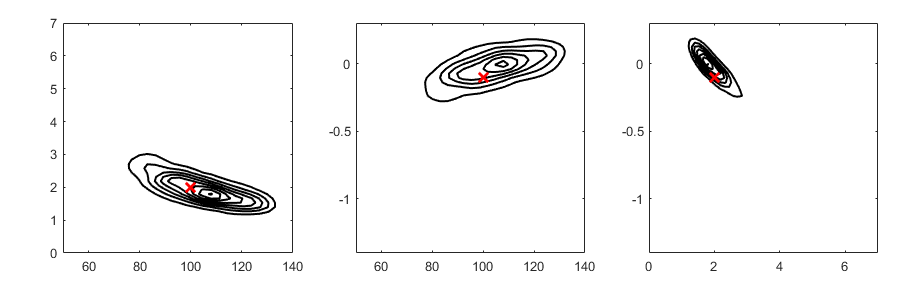}\label{figsub:results_toad_bivariate_simulated_bsl}}
	\caption{Contour plots of the approximate bivariate posteriors based on various methods for the simulated data of the stereological extremes example. True parameter values are shown as red crosses.}
	\label{fig:results_stereo_bivariate_simulated}
\end{figure}

\begin{figure}[H]
	\centering
	\subfigure[CvM]{\includegraphics[scale=0.35]{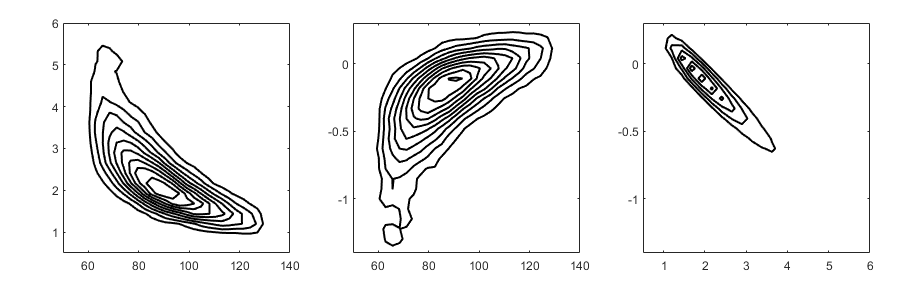}\label{figsub:results_toad_bivariate_real_cvm}}
	\subfigure[Wass]{\includegraphics[scale=0.35]{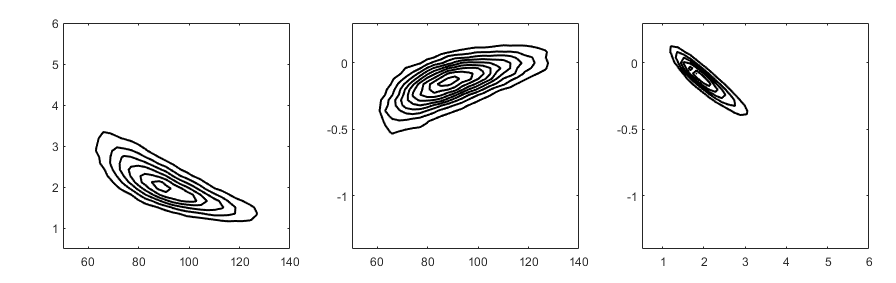}\label{figsub:results_toad_bivariate_simulated_wass}}
	\subfigure[MMD]{\includegraphics[scale=0.35]{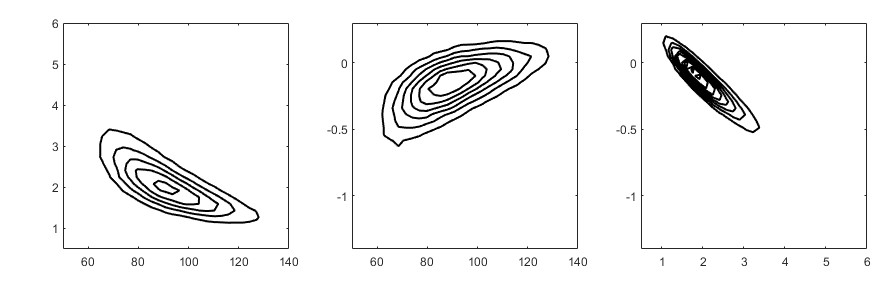}\label{figsub:results_toad_bivariate_simulated_mmd}}
	\subfigure[ABC 4stats]{\includegraphics[scale=0.35]{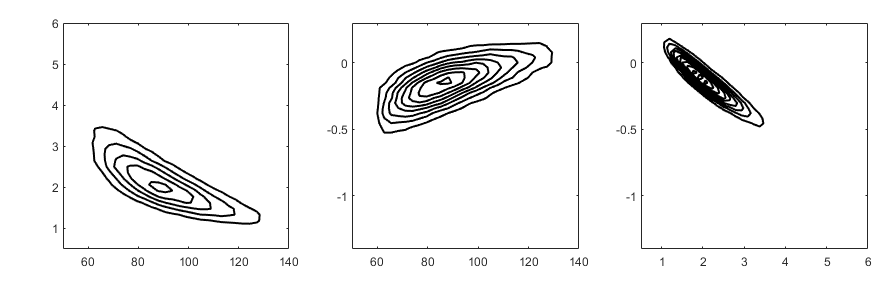}\label{figsub:results_stereo_bivariate_real_abc_4stats}}
	\subfigure[ABC Summ]{\includegraphics[scale=0.35]{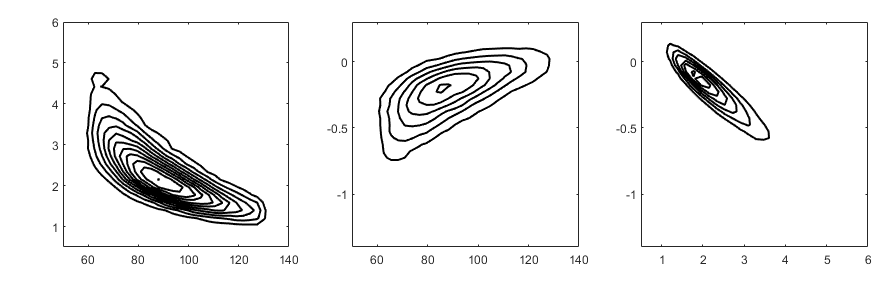}\label{figsub:results_toad_bivariate_simulated_quantiles}}
	\subfigure[KDE]{\includegraphics[scale=0.35]{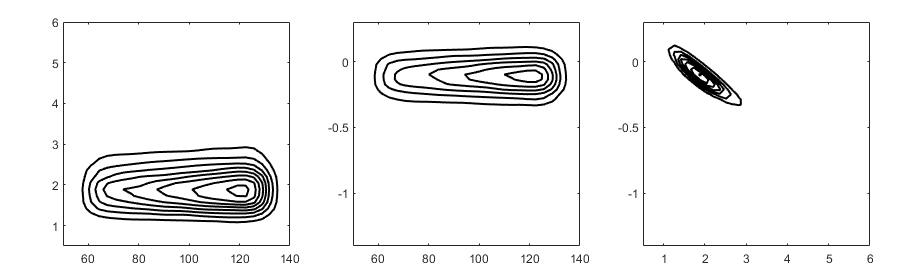}\label{figsub:results_stereo_bivariate_real_abc_aux}}
	\subfigure[BSL]{\includegraphics[scale=0.35]{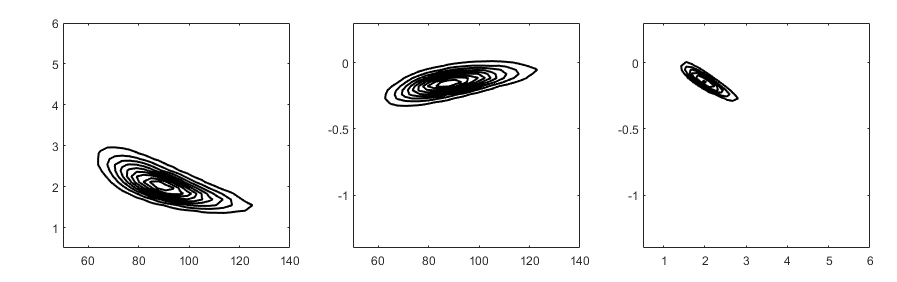}\label{figsub:results_toad_bivariate_simulated_bsl}}
	\caption{Contour plots of the approximate bivariate posteriors based on various methods for the real data of the stereological extremes example. }
	\label{fig:results_stereo_bivariate_real}
\end{figure}

The estimated bivariate posterior distributions for the simulated and real data are shown in Figures \ref{fig:results_stereo_bivariate_simulated} and \ref{fig:results_stereo_bivariate_real}, respectively.  Please see Section 4.3 in the main paper for further details, including the abbreviations used within the following figures.

\section*{Appendix D: Additional Details and Results for Toad Example}

The pseudocode for simulating the random return toad model can be found in Algorithm \ref{alg:toad}.  The simulation produces a matrix $\mathbf{X}$ of dimension $n_d \times n_t$, which effectively holds location data based on distance travelled each day.  For day $t$, toad $j$ is assumed to return to one of the previous locations with a probability $p_0$. If this occurs, then we set $\mathbf{X}_{t,j} = \mathbf{X}_{\mathrm{loc},j}$ where loc $\sim \mbox{discrete}\mhyphen\mathcal{U}(1,t-1)$.  If a toad does not return to a previously observed location, then we add a displacement $\delta \sim \alpha\mhyphen\mathrm{stable}(\alpha, \xi)$ to the location from day $t-1$, $\mathbf{X}_{t,j} = \mathbf{X}_{t-1,j} + \delta$, which effectively creates a new location.  

\begin{algorithm}[tbh]
	\caption{{Simulation of random return toad model} \label{alg:toad} 
		\vspace{2mm}
		\newline
		{\em Inputs:} 
		Parameter $\theta = (\alpha,\xi,p_0)^\top$, number of days to simulate $n_d$ and number of toads being tracked $n_t$.
		\vspace{2mm}
		\newline
		{\em Outputs:} $\mathbf{X}$ as a matrix of size $n_d \times n_t$
	}
	\begin{algorithmic}[1]
		\State Set $\mathbf{X}$ as a matrix of zeros of size $n_d \times n_t$
		\For{$i=2$ to $n_d$} \Comment{loop over days}
		\For{$j=1$ to $n_t$} \Comment{loop over toads}
		\If{$\mathcal{U}(0,1) > p_0$}
		\Comment{toad does not return, stays at new location}
		\State $\delta \sim \alpha\mhyphen\mathrm{stable}(\alpha, \xi)$
		\State $\mathbf{X}_{i,j} = \mathbf{X}_{i-1,j} + \delta$
		\Else
		\Comment{toad returns to one of the previous refuge sites}
		\State loc $ \sim \mbox{discrete}\mhyphen\mathcal{U}(1,i-1)$
		\State $\mathbf{X}_{i,j} = \mathbf{X}_{\mathrm{loc},j}$
		\EndIf
		\EndFor
		\EndFor
	\end{algorithmic}
\end{algorithm}

An initial summarisation is performed that involves computing the relative moving distances for time lags of $1,2,4,8$ days. For instance, $\mathbf{x}_1$ consists of the relative displacement information at a lag of one day, $\mathbf{x}_1 = \{|\Delta x| = |\mathbf{X}_{i,j}-\mathbf{X}_{i+1,j}| ; 1 \leq i \leq n_d-1, 1 \leq j \leq n_t \}$.  For each lag, we split the displacement vector into two sets.  The first set holds displacements where $|\Delta x| < 10$m, and these are taken as returns (the toad has returned to same location in $k$ days where $k$ is the time lag), and we simply record the number of returns.  The second set holds the vector of displacements that are greater than 10m (non-returns).  Combining these two datasets (the number of returns and the vector of non-return displacements) for each of the four separate lags (1, 2, 4 and 8 days) produces the data used for the analysis. 


For the summary statistic based approaches we use the following data summarisation.  For a given time lag, we use the following summary statistics for the non-returns data.  We firstly compute the $0,0.1,\dots,1$ quantiles, and then we use the log of the differences of the adjacent quantiles as summary statistics; the median of the non-returns data is also used as a summary statistic.  We also append the number of returns as a summary statistic.  This produces 12 summary statistics for each lag, and given that there are 4 different lag times, there are 48 summary statistics in total. 

For the real data there are various missing values in the data matrix $\boldsymbol{Y}$.  After computing the initial summarisation $\mathbf{y}$ as described above, the missing values in $\mathbf{y}$ are removed before computing the summary statistics.  When the real data is analysed, after each model simulation we place missing values in $\mathbf{X}$ in the same positions as they appear in $\boldsymbol{Y}$.

\begin{figure}[h]
	\centering
	\subfigure[CvM]{\includegraphics[scale=0.7]{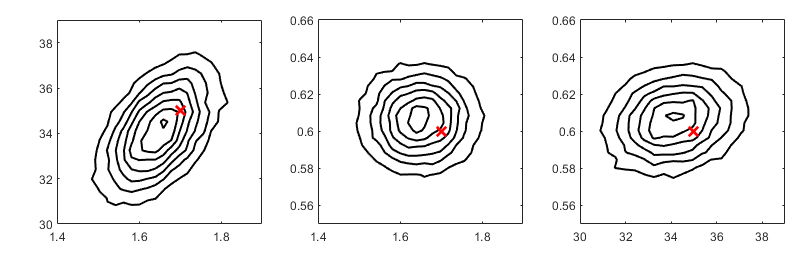}\label{figsub:results_toad_bivariate_simulated_cvm}}
	\subfigure[Wass log]{\includegraphics[scale=0.7]{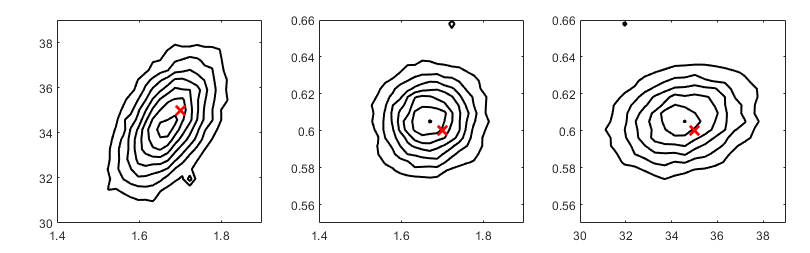}\label{figsub:results_toad_bivariate_simulated_wass_log}}
	\subfigure[ABC]{\includegraphics[scale=0.7]{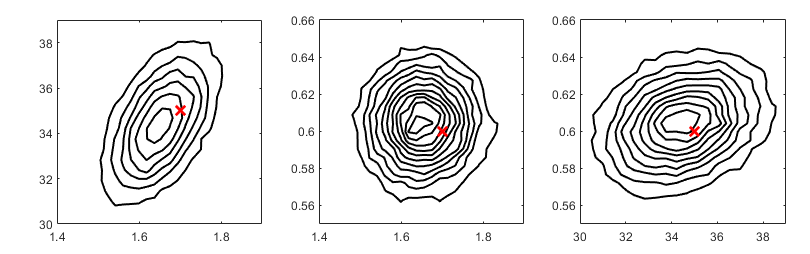}\label{figsub:results_toad_bivariate_simulated_abc}}
	\subfigure[BSL]{\includegraphics[scale=0.7]{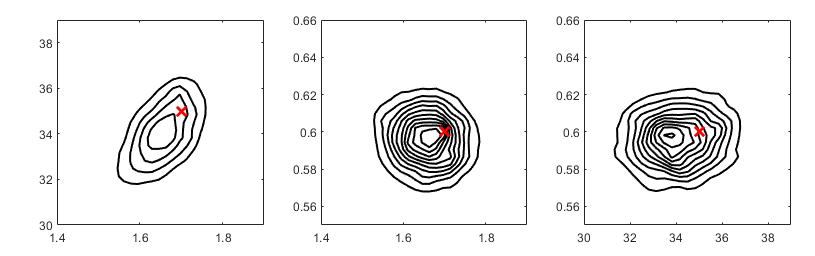}\label{figsub:results_toad_bivariate_simulated_bsl}}
	\caption{Contour plots of the approximate bivariate posteriors based on various methods for the simulated data of the toad example. True parameter values are shown as red crosses.}
	\label{fig:results_toad_bivariate_simulated}
\end{figure}

\begin{figure}[h]
	\centering
	\subfigure[CvM]{\includegraphics[scale=0.55]{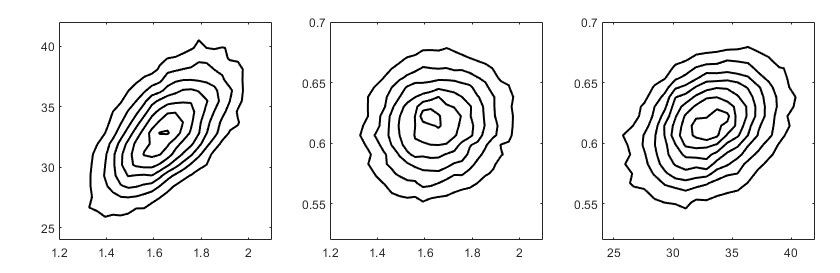}\label{figsub:results_toad_bivariate_real_cvm}}
	\subfigure[Wass (log)]{\includegraphics[scale=0.55]{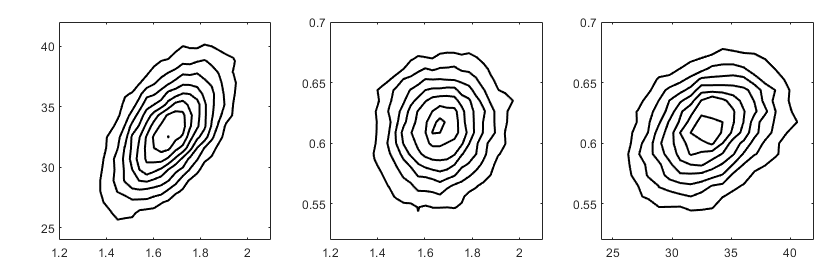}\label{figsub:results_toad_bivariate_real_wass_log}}
	\subfigure[MMD (log) ]{\includegraphics[scale=0.55]{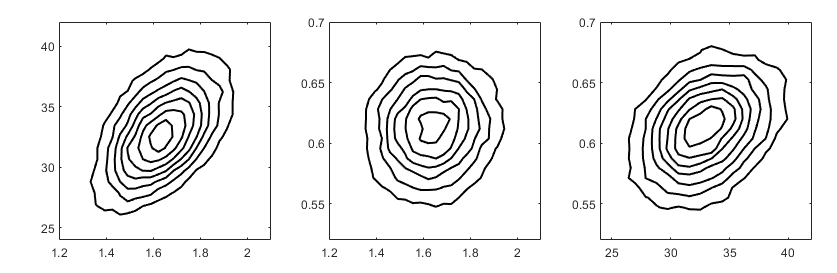}\label{figsub:results_toad_bivariate_real_mmd_log}}
	\subfigure[ABC]{\includegraphics[scale=0.55]{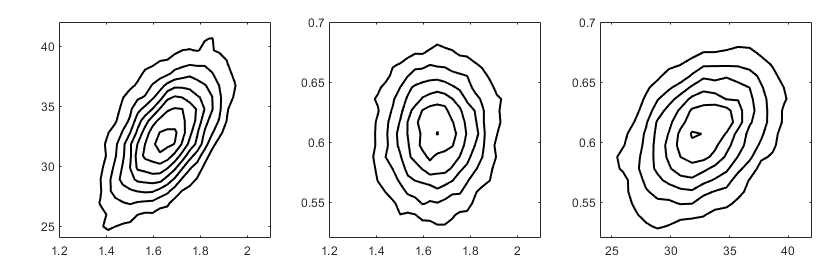}\label{figsub:results_toad_bivariate_real_quantiles}}
	\subfigure[BSL]{\includegraphics[scale=0.55]{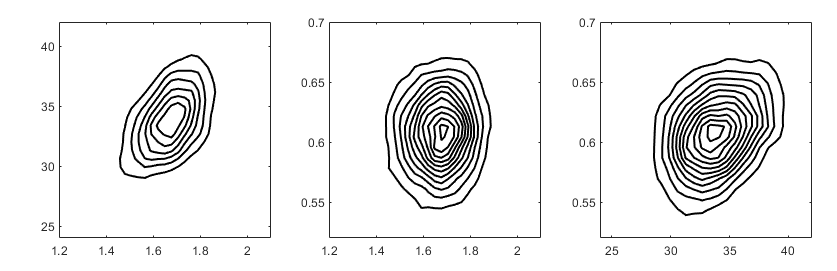}\label{figsub:results_toad_bivariate_real_bsl}}
	\caption{Contour plots of the approximate bivariate posteriors based on various methods for the real data of the toad example.}
	\label{fig:results_toad_bivariate_real}
\end{figure}

Figures \ref{fig:results_toad_bivariate_simulated} and \ref{fig:results_toad_bivariate_real} show approximate bivariate posteriors obtained from various methods for the simulated and real data, respectively.  Please see Section 4.4 in the main paper for further details, including the abbreviations used within the following figures.

\section*{Appendix E: Additional Results for Toggle Switch Example}\label{appn} 
\begin{figure}[h]
	\centering
	\subfigure[2nd dataset]{\includegraphics[scale=0.3]{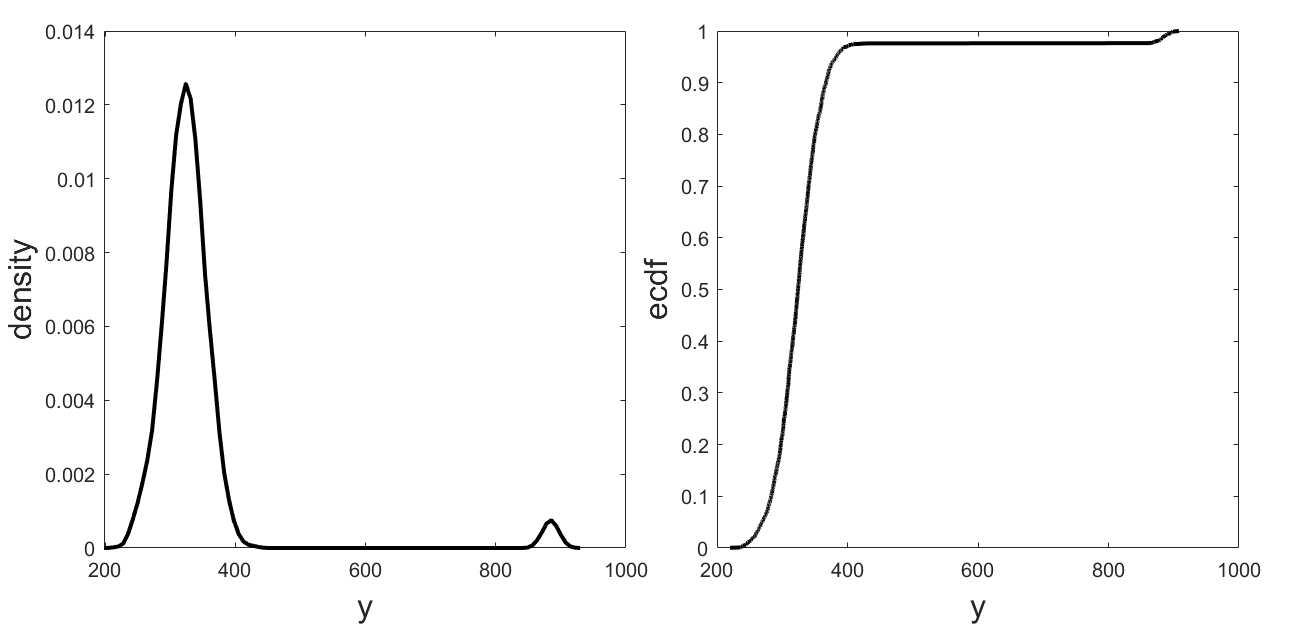}\label{figsub:data_toggle_d2}}
	\subfigure[3rd dataset]{\includegraphics[scale=0.3]{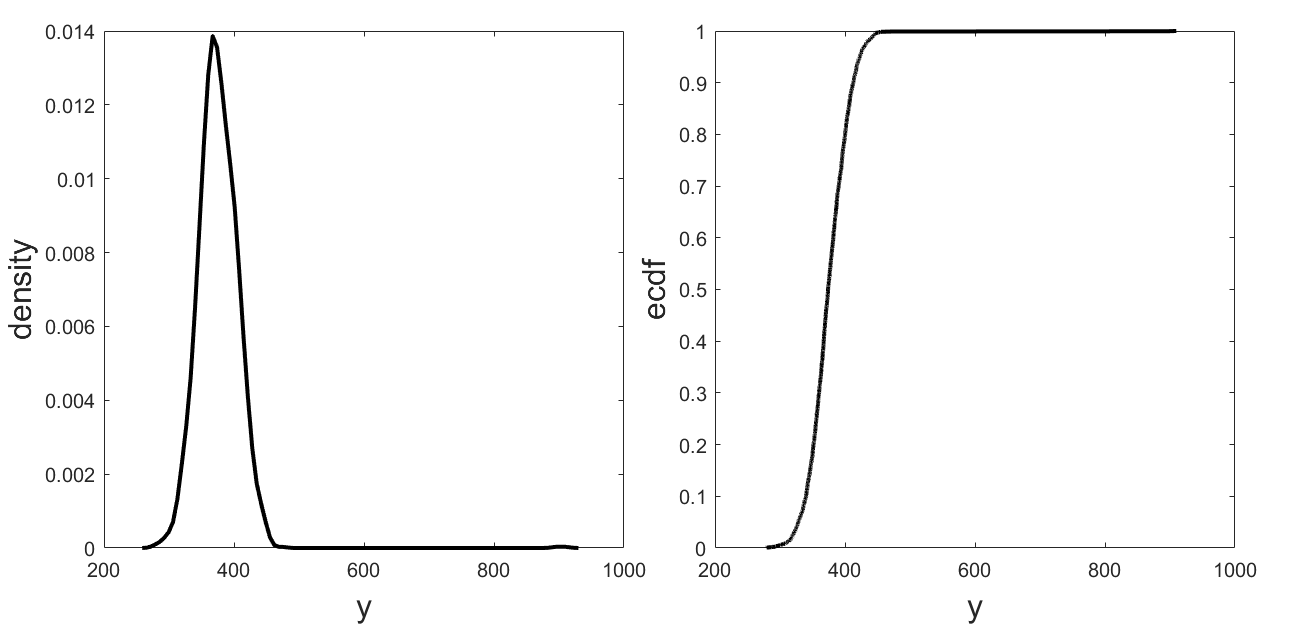}\label{figsub:data_toggle_d3}}
	\subfigure[4th dataset]{\includegraphics[scale=0.3]{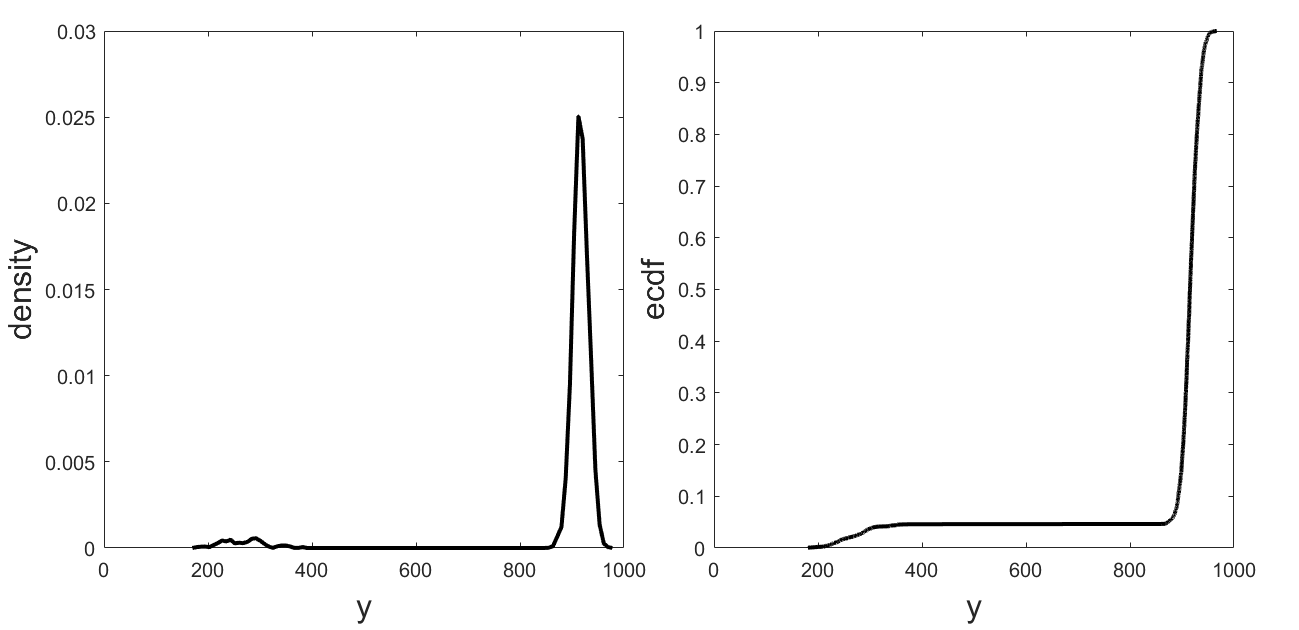}\label{figsub:data_toggle_d4}}
	\subfigure[5th dataset]{\includegraphics[scale=0.28]{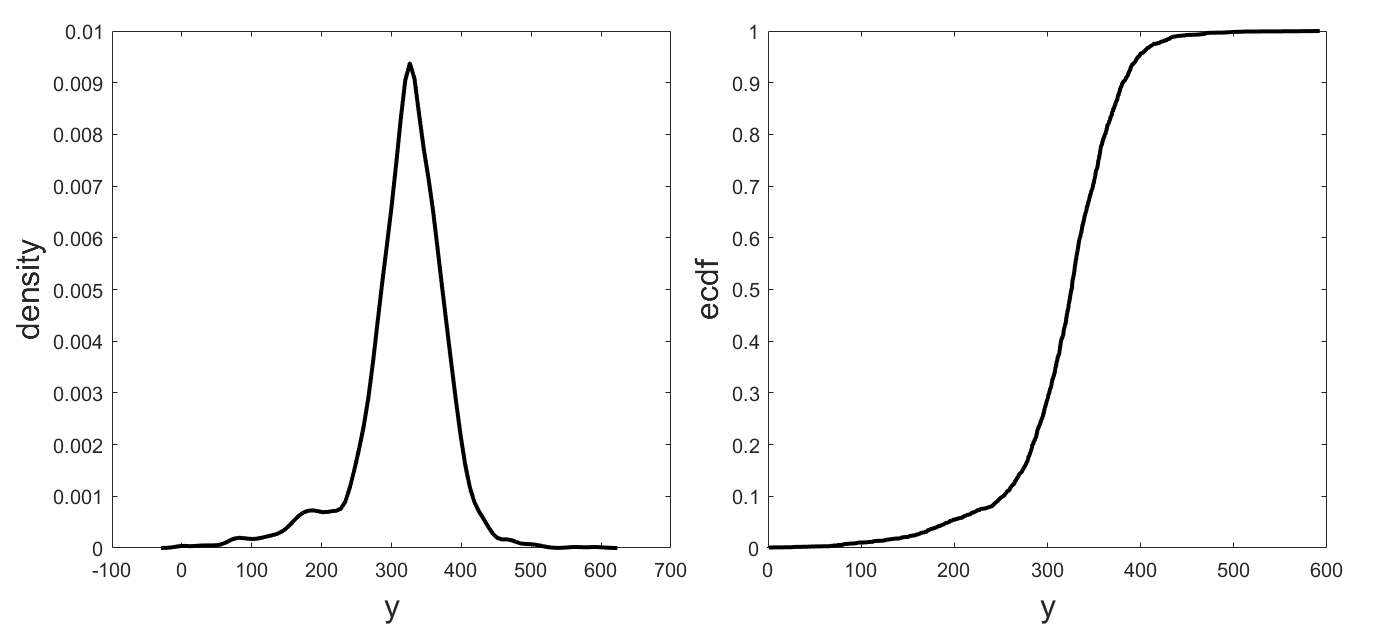}\label{figsub:data_toggle_d5}}
	\caption{Additional simulated datasets for the toggle switch example.}
	\label{fig:data_toggle_extra}
\end{figure}

\begin{figure}[h]
	\centering
	\includegraphics[scale=0.70]{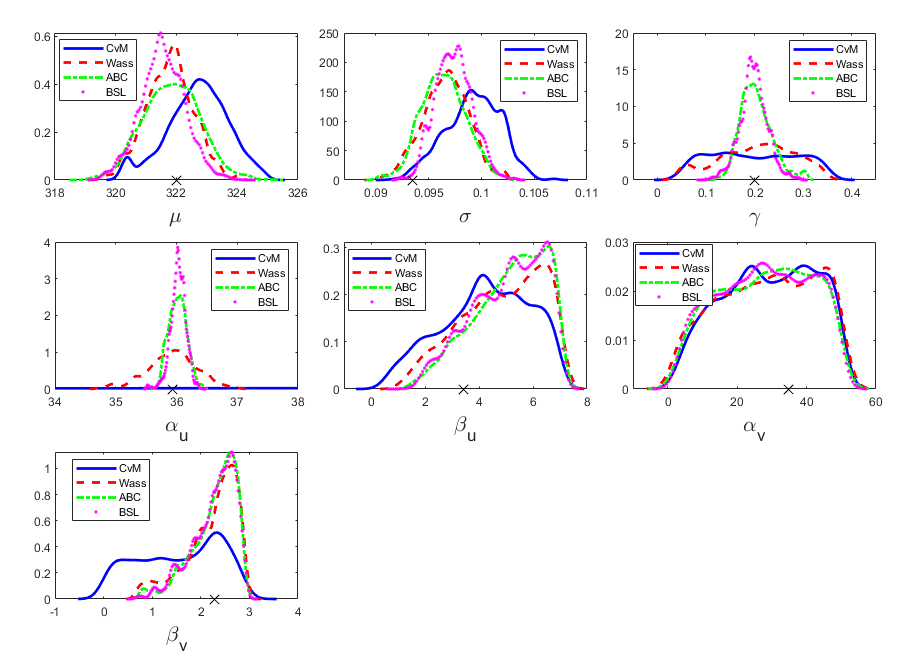}
	\caption{Comparison of estimates of the univariate likelihood-free posterior distributions for the toggle switch example based on a second simulated dataset. }
	\label{fig:results_toggle_d2}
\end{figure}

\begin{figure}[h]
	\centering
	\includegraphics[scale=0.70]{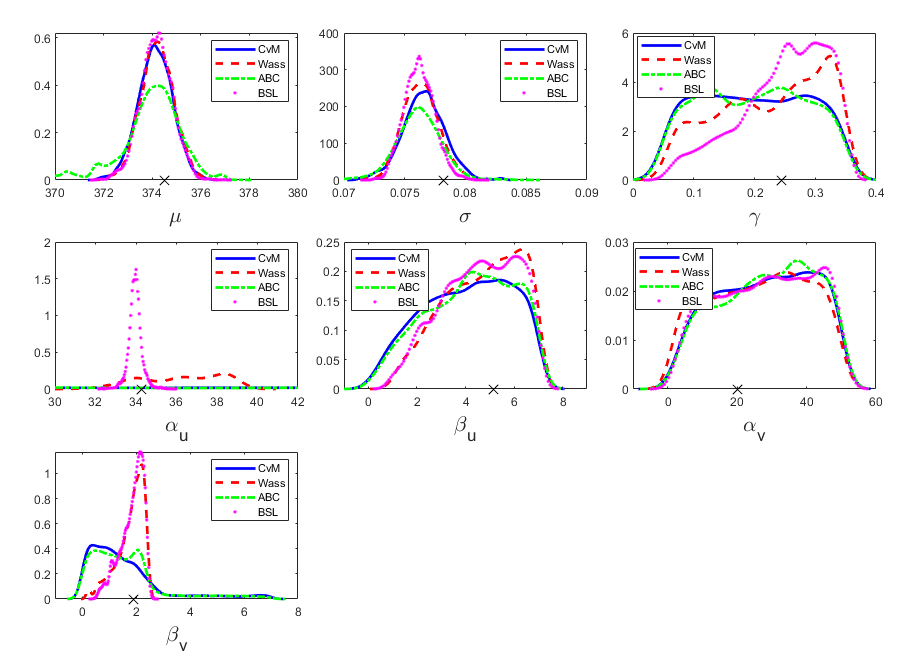}
	\caption{Comparison of estimates of the univariate likelihood-free posterior distributions for the toggle switch example based on a third simulated dataset. }
	\label{fig:results_toggle_d3}
\end{figure}

\begin{figure}[h]
	\centering
	\includegraphics[scale=0.70]{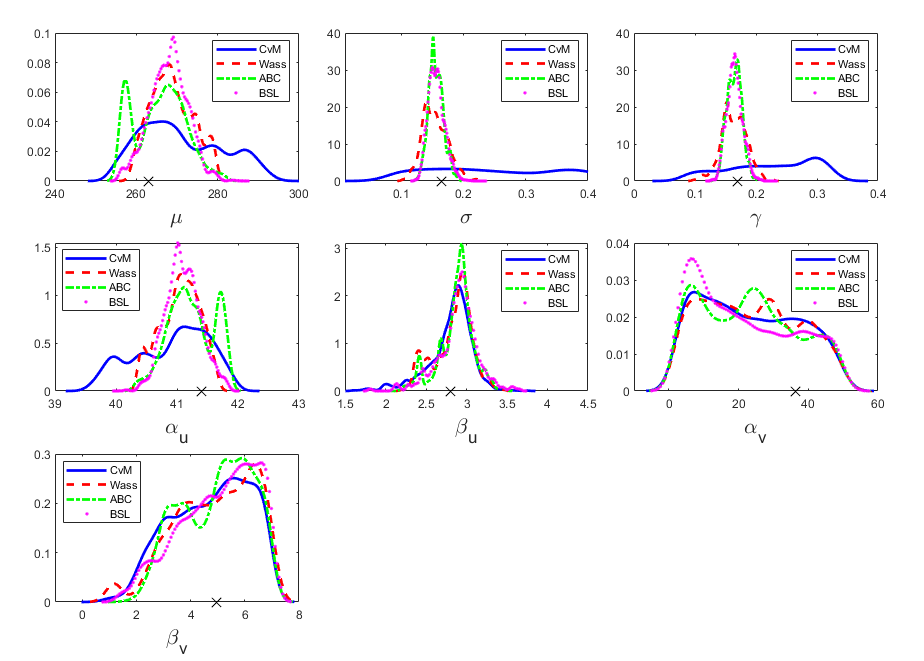}
	\caption{Comparison of estimates of the univariate likelihood-free posterior distributions for the toggle switch example based on a fourth simulated dataset. }
	\label{fig:results_toggle_d4}
\end{figure}

\begin{figure}[h]
	\centering
	\includegraphics[scale=0.70]{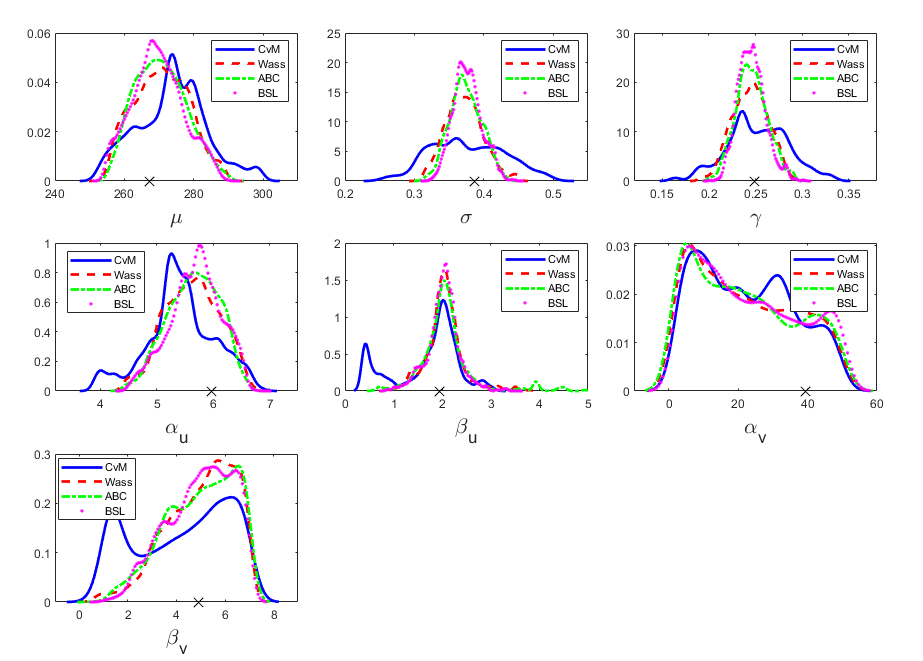}
	\caption{Comparison of estimates of the univariate likelihood-free posterior distributions for the toggle switch example based on a fifth simulated dataset. }
	\label{fig:results_toggle_d5}
\end{figure}

Here we show more results for the toggle switch example for some datasets generated from the prior predictive distribution.  The datasets are shown in Figure \ref{fig:data_toggle_extra} and the results are shown in Figures \ref{fig:results_toggle_d2}-\ref{fig:results_toggle_d5}.  Please see Section 4.5 of the main paper for further details.

h

\end{document}